\documentclass[apj,twocolumn]{emulateapj_mod}
\usepackage{epsfig,apjfonts,mathptmx}

\def\gtsima{$\; \buildrel > \over \sim \;$}
\def\ltsima{$\; \buildrel < \over \sim \;$}
\def\prosima{$\; \buildrel \propto \over \sim \;$}
\def\gsim{\lower.5ex\hbox{\gtsima}}
\def\lsim{\lower.5ex\hbox{\ltsima}}
\def\simgt{\lower.5ex\hbox{\gtsima}}
\def\simlt{\lower.5ex\hbox{\ltsima}}
\def\simpr{\lower.5ex\hbox{\prosima}}

\def\h1{$h^{-1}$}
\def\eeq{\end{equation}}
\def\beq{\begin{equation}}

\shorttitle{Joint selection of $1.4\simlt z\simlt2.5$ star-forming and passive galaxies}
\shortauthors{E. Daddi et al.}

\begin{document}

\title{
A new photometric technique for the joint selection of star-forming\\
and passive galaxies at {$1.4\simlt \lowercase{z}\simlt2.5$}$^1$
}

\author{E. Daddi\altaffilmark{2},
	A. Cimatti\altaffilmark{3},
	A. Renzini\altaffilmark{2},
	A. Fontana\altaffilmark{4},
	M. Mignoli\altaffilmark{5},
	L. Pozzetti\altaffilmark{5},
	P. Tozzi\altaffilmark{6},
	G. Zamorani\altaffilmark{5}
}

\altaffiltext{1}{Based on observations collected at the European
Southern Observatory, Chile (ESO programs 70.A-0140, 70.A-0548,
168.A-0485, 170.A-0788), and with the NASA/ESA {\em
Hubble Space Telescope}, 
which is operated by AURA Inc, under NASA contract NAS 5-26555.}  
\affil{$^2$European Southern Observatory,
Karl-Schwarzschild-Str. 2, D-85748 Garching, Germany
} 
\affil{$^3$INAF--Osservatorio Astrofisico di
Arcetri, L.go E. Fermi 5, Firenze, Italy} 
\affil{$^4$INAF--Osservatorio Astronomico di Roma, via Dell'Osservatorio 2, Monteporzio, Italy}
\affil{$^5$INAF--Osservatorio Astronomico di Bologna, via
Ranzani 1, Bologna, Italy} 
\affil{$^6$INAF--Osservatorio
Astronomico di Trieste, via Tiepolo 11, Trieste, Italy}

\begin{abstract}
A simple two color selection based on $B$-, $z$-, and $K$- band
photometry is proposed for culling galaxies at $1.4\simlt
z\simlt2.5$ in $K$-selected samples and classifying them as
star-forming or passive systems. The method is calibrated on the
highly complete spectroscopic redshift database of the K20 survey,
verified with simulations and tested on other datasets. Requiring 
$BzK=(z-K)_{AB}-(B-z)_{AB}>-0.2$ allows to select actively star-forming 
galaxies at $z\simgt1.4$, independently on their dust reddening. Instead,
objects with $BzK<-0.2$ and $(z-K)_{AB}>2.5$ colors include passively 
evolving galaxies at $z\simgt1.4$, often with spheroidal morphologies. 
Simple recipes 
to estimate the reddening, SFRs and masses of $BzK$-selected galaxies
are derived, and calibrated on $K<20$ galaxies.
These $K<20$ galaxies have
typical stellar masses $\sim10^{11}M_\odot$, and sky and volume density
of $\sim1$ arcmin$^{-2}$ and $\sim10^{-4}$Mpc$^{-3}$ respectively.
Based on their UV (reddening-corrected), X-ray and radio luminosities, 
the $BzK$--selected star-forming galaxies with $K<20$ turn out to have average 
$SFR\approx 200\ M_{\odot} yr^{-1}$, and median reddening
$E(B-V)\sim0.4$. 
This SFR is a factor of 10 higher than that 
of $z\sim1$ dusty EROs, and a factor of 3 higher than found for 
$z\sim2$ UV selected galaxies, both at similar $K$ limits. 
Besides missing the passively evolving galaxies, 
the UV selection appears to miss some relevant fraction of the $z\sim2$
star-forming galaxies with $K<20$, 
and hence of the (obscured) star-formation rate density at this redshift.
The high SFRs and masses add to other existing evidence that these $z=2$
star-forming galaxies may be among the precursors of $z=0$ early-type galaxies.
A V/V$_{max}$ test suggests that such a population 
may be increasing in number density with increasing redshift. 
Theoretical models cannot reproduce simultaneously the space density
of both passively evolving and highly star-forming galaxies at $z=2$.
In view of Spitzer Space Telescope observations, an analogous
technique based on
the $RJL$ photometry is proposed to complement the $BzK$ selection
and to identify massive galaxies at $2.5\simlt z\simlt 4.0$.
By selecting passively evolving galaxies as well as actively
star-forming galaxies (including strongly dust reddened ones), 
these color criteria should help in completing the census of the stellar
mass and of the star-formation rate density at high redshift.
\end{abstract}

\section{Introduction}

Tracing and understanding the history of cosmic star formation and the 
growth of the cosmic stellar mass density are currently the objects
of major observational efforts (e.g., Dickinson et al. 2003; Fontana et al.
2003; Rudnick et al. 2003), yet much remains to be done before
reaching fully satisfactorily conclusions.  Obtaining a complete
census of the populations of high-redshift galaxies, and their
physical characterization (stellar mass and star-formation rate, SFR)
is necessary to observationally map the processes that lead to galaxy
formation and evolution. This requires direct spectroscopic
identification (e.g., Fontana et al. 2004; Glazebrook et al. 2004),
but building large samples of spectroscopically confirmed galaxies at
high redshift is a time consuming process. This is especially true for
magnitude-limited samples, as high-$z$ galaxies represent a small
fraction of the galaxy counts at faint magnitudes in the optical and
infrared bands. Therefore, techniques for pre-selecting high redshift
targets are required to focus the spectroscopic multiplex capability
on the interesting objects. Photometric redshifts from deep multicolor
datasets offer an alternative to massive spectroscopic efforts
(e.g. Bolzonella et al. 2002; Firth et al. 2002; Poli et al. 2003), 
but the results may suffer from biases that are
difficult to quantify, and their accuracy depends critically on the
quality of the photometry and the colors and redshifts of the objects.

An efficient alternative to either photometric redshifts or magnitude
limited surveys is offered by simpler single or two-color criteria, then
followed by targeted spectroscopy. The best known example is the
dropout technique for selecting Lyman-Break Galaxies (LBG, Steidel et
al. 1996; 2003) from their $U_{\rm n}GR_{\rm s}$ rest-frame UV colors, which
opened the research field on normal star-forming galaxies at
$z\sim3$. Recently, the same technique has been extended to work at
lower redshifts $1.4<z<2.5$ (Erb et al. 2003; Adelberger et al. 2004)
and large samples of UV-selected objects have been spectroscopically
confirmed at $z\sim2$ (Steidel et al. 2004).

Star forming galaxies can be selected as LBGs only if they are UV
bright (i.e.  actively star forming) and not heavily reddened by dust.
Currently, the best possible alternatives to find dust-enshrouded high
redshift star-forming objects include detecting them from their far-IR or
sub-mm emission due to cold dust (Franceschini et al. 2001; Smail et
al. 2002; Champan et al. 2003), observing the emission at
X-ray or radio wavelengths that are not extincted by dust (e.g.,
Norman et al. 2004; Haarsma et al. 2000), and selecting very red 
objects in near-infrared samples that are less affected by dust extinction
(Cimatti et al. 2002a, 2003).

Besides targeting star-forming galaxies, color criteria have also been
used to search for passively evolving galaxies at high redshifts.  A
simple criterion is the one used for Extremely Red Objects (ERO),
selected according to their very red optical to near-IR colors,
e.g. $R-K>5$ or $I-K>4$ (Thompson et al 1999; Daddi et al. 2000a;
2000b; Firth et al. 2002; Roche et al. 2002; 2003; Miyazaki et
al. 2003; McCarthy 2004). Spectroscopy showed that EROs include both
old passive galaxies at $0.8<z<2$ and dusty star-forming systems at
similar redshifts (Cimatti et al. 2002a, 2003; Yan et al. 2004). 
The bulk of EROs, however, is made of galaxies at redshifts 
$z\sim 1$ with only a small fraction being at $z\simgt2$, a crucial 
redshift range for the evolution of galaxies.  In order to identify 
old stellar systems at $z\sim2$ or beyond, color criteria based 
on $J$ and $K$ imaging have been proposed (e.g., Pozzetti \& 
Mannucci 2000; Totani et al. 2001; Franx et al. 2003; Saracco et al. 2004). 
As confirmed by van Dokkum et al. 
(2003), objects with red spectral energy distributions at $z\simgt2$, 
can be 
selected requiring very red colors $J-K>2.3$ (Vega scale).

Although the selection of objects with extremely red colors has been
quite successful, one could expect that moderately old, or moderately
reddened, objects exist at high redshift that would be missed by both
the ``red-color" techniques and by the UV techniques.  In addition, it
appears unsatisfactory to use so many different color criteria in order
to build representative samples of galaxies as a function of
redshifts, as the physical relations between these different classes
remain unknown, and the selection biases not fully understood.

The recently completed K20 survey (Cimatti et al. 2002a,b,c;
Daddi et al. 2002; Pozzetti et al. 2003; Fontana et al. 2004) 
has reached an unprecedented  spectroscopic completeness ($>92\%$) for a
sample of $K<20$ galaxies. In particular, for one of the two K20 fields
(included in the CDF-S/GOODS-S field) the recent deep
observations (Daddi et al. 2004, hereafter D04; Cimatti et al. 2003, 2004;
Vanzella et al. 2004\footnote{Publicly available ESO observations 
obtained as part of the GOODS project: http://www.eso.org/science/goods/}) 
allowed to reach a spectroscopic completeness of $>94\%$. This includes the 
presence of a significant fraction of high redshift, $z>1.4$, galaxies
(redshift desert coverage).

In this paper we take advantage of the K20 spectroscopic sample to
define a simple two-color criterion based on the $B$-, $z$- and $K$-band
photometry which, with a minimal contamination from lower redshift
galaxies, is capable of identifying the full range of high-redshift
$z>1.4$ galaxies in our $K$-selected sample, including both actively 
star-forming (D04) and old passive objects (Cimatti et al. 2004), 
and to distinguish between the two classes.

For galaxies identified with this criterion (at least to $K=20$) 
it is found that a $K$-band selection
is close to a galaxy stellar-mass selection, while
a $K$-selected sample of star-forming galaxies allows to reach completeness
down to a given star-formation rate limit almost independently of dust
reddening. Therefore, the technique offers 
a powerful tool to explore with the minimum possible biases the 
histories of cosmic star-formation  and cosmic stellar-mass build-up 
at $z\sim2$.
We discuss in detail to which extent the cosmic stellar-mass and star-formation
rate density can be estimated with the properties of galaxies in the proposed
two-color diagram. The X-ray  and radio properties of
$K$-selected star-forming galaxies are also investigated
in order to provide an independent estimate
of their star-formation rates.

The paper is organized as follow. The spectroscopic and imaging datasets
used in the paper are described in Section~\ref{sec:data}. The $BzK$
selection and classification
technique for $1.4<z<2.5$ galaxies is empirically calibrated in
Section~\ref{sect:BzK}, checked against stellar population models in
Section~\ref{sec:modeling} and compared to HST morphological classification in 
Sect.~\ref{sect:HST}. The SFR and mass content of $z\sim2$ galaxies 
are described in Section~\ref{sec:SFR} and \ref{sec:masses}, together with
methods to obtain ensemble averages from the $BzK$ photometry alone.
Section~\ref{sec:other} compares the samples selected with the $BzK$
technique to those of other criteria, including UV selected $z\sim2$ galaxies,
EROs and $J-K$ red galaxies.
We extend the technique for use at higher
redshifts using Spitzer Space Telescope (SST) imaging in
Section~\ref{sec:SIRTF}. The results are discussed in
Section~\ref{sec:discussion} and summary and conclusions are in
Section~\ref{sec:summary}.

We use the Salpeter IMF extending between 0.1 and 100 $M_\odot$
and a WMAP cosmology with $\Omega_\Lambda, \Omega_M = 0.73, 0.27$, and
$h = H_0$[km s$^{-1}$ Mpc$^{-1}$]$/100=0.71$.

\section{The data}
\label{sec:data}

\subsection{The K20/GOODS Field}

The K20 survey has obtained spectra for $545$ objects selected in the
$K$-band over
two widely separated fields for a total area of 52 arcmin$^2$, including
a 32 arcmin$^2$ region of the GOODS-South field (Cimatti et al. 2002b). 
Of the 347 objects with $K<20$ in this area, 328 have been spectroscopically 
identified at the moment by complementing the K20 spectroscopy with a
few additional redshifts from the ESO/GOODS public spectroscopy
(Vanzella et al.\ 2004). The identified targets include
292 extragalactic objects and 36 stars, while the
residual 19 objects have only photometric redshifts. Among these 311
galaxies, 19 have $z_{\rm spec}>1.4$ (6\% of the sample) and 13 have
$z_{\rm phot}>1.4$, or $\sim 10\%$ lie at an estimated redshift beyond
1.4. As already pointed out in previous K20 papers (Cimatti et al. 2002c; 
D04), no K20 galaxies were expected to lie at these high redshifts
($z\sim2$) based on current semi-analytical models of galaxy formation. 

\begin{figure}[ht]
\centering 
\includegraphics[width=8.8cm]{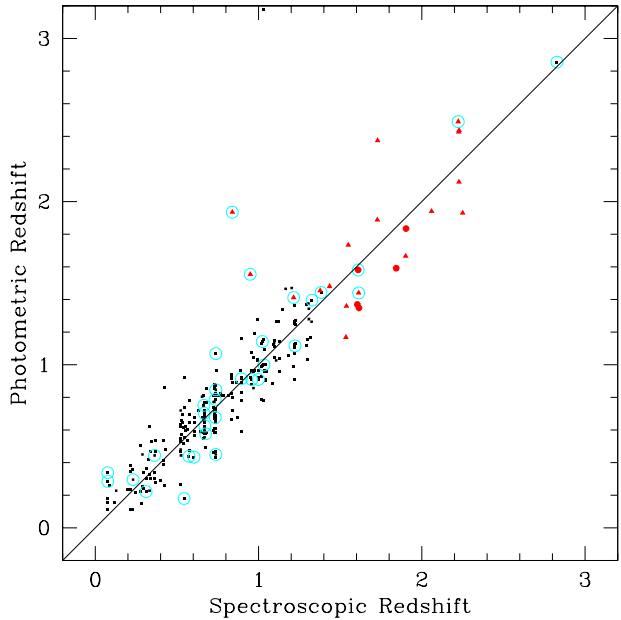}
\caption{Comparison of spectroscopic and photometric redshifts for the
311 extragalactic objects in the K20/GOODS area.
Filled triangles show objects having $BzK>-0.2$ (Eq.~\ref{eq:BzK}), 
filled circles are old galaxies at $z>1.4$, circled symbols are X-ray 
detected sources and small symbols are $z<1.4$ (or $z>2.5$)
galaxies. The photometric 
redshifts of old $z>1.4$ galaxies are slightly but systematically 
underestimated, 
probably because the Coleman et al. (1980) templates
are too red with respect to these $\sim1$--2 Gyr old passive galaxies.
Using the full library of BC03 SSP models with no dust reddening,
the photometric redshifts of old $z>1.4$ galaxies become much more
accurate and without systematic effects. 
}
\label{fig:BzK_photoz}
\end{figure}

In addition to spectroscopy, deep and high quality imaging and
photometry is available for this field, including
ground-based $BVRIzJHK$ imaging with very good seeing (generally 0.4--0.7$''$)
obtained with FORS1, FORS2 and ISAAC 
at the VLT  (the same imaging dataset used in D04), together with the 
HST+ACS $bviz$ data released by the GOODS Team (Giavalisco et al. 2004). 
Shallower $U$-band imaging from Arnouts et al. (2001) was also used,
obtained at the ESO 2.2m telescope.
Photometric redshifts were computed for all galaxies with {\em hyperz} 
(Bolzonella et al. 2000) using the available multiwavelength 
photometry. These are updated with respect to D04 and Cimatti et al. 
2002 because of the inclusion of the final ACS photometry ($bviz$) 
from GOODS. Fig.~\ref{fig:BzK_photoz} shows the comparison of these 
photometric vs. the spectroscopic redshifts.

\begin{figure}[ht]
\centering 
\includegraphics[width=8.8cm]{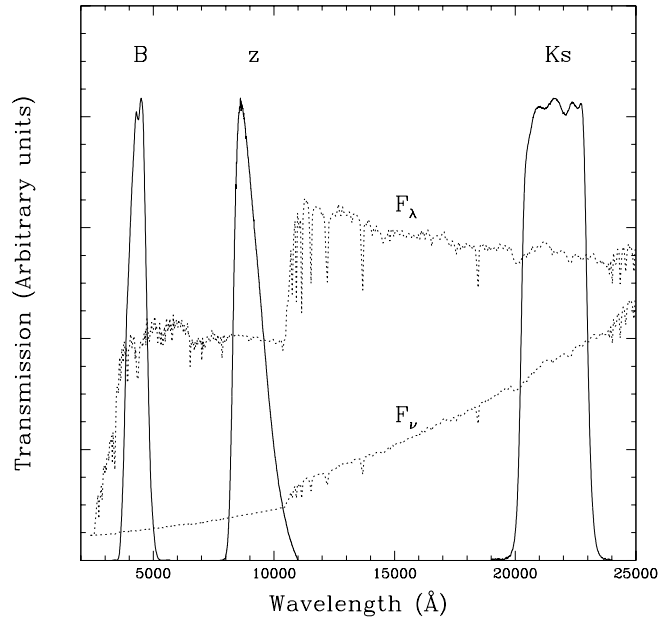}
\caption{The total transmission curves (including the detectors QE
and atmospheric transmission) of the $BzK$ filters used to define
the criteria for selecting $z>1.4$ galaxies. For the $z$-band filter,
the HST+ACS curve is shown, with
the VLT filter being very similar and only slightly less extended to the red. 
The best fit BC03 model to the SED of a galaxy at $z=1.729$ is 
shown for reference, on linear flux scale both for $F_\nu$ and $F_\lambda$.
}
\label{fig:pfilter}
\end{figure}

The $B$-, $z$- and $K$-band photometry, on which most of the paper 
is focused, is 
based on the Bessel $B$-band, F850LP $z$-band and the $K_{\rm s}$-band,
(referred to as the $K$ band in the rest of this paper). The F850LP
zeropoint was rescaled to match the photometry of the $z$-band
imaging based on VLT+FORS1 Gunn-$z$ filter, that is considerably less deep 
than the GOODS $z$-band imaging. No color term was considered given
the overall similarity of the two $z$-band filters whose effective
wavelengths differ only by about 1\%.
Our ground based VLT Bessel $B$-band imaging is instead 
significantly deeper than the GOODS ACS imaging with F435W when
measuring on apertures $\simgt1"$ comparable to the size of the K20
galaxies. The $K$-band data was obtained with VLT+ISAAC.
Fig.~\ref{fig:pfilter} shows the total efficiency of the  $BzK$
photometric systems as a function of wavelength. In order to allow
a fine tuning of the photometry of objects from surveys using 
slightly different $BzK$ filter sets we make publicly
available\footnote{http://www.arcetri.astro.it/$\sim$k20/releases/} 
the $B-z$ and $z-K$ magnitudes of the stars identified in the GOODS area
of the K20 survey. 
By matching the colors of the stellar sequence to the K20 one, it is 
possible to accurately apply the selection criteria described
in Section~\ref{sect:BzK}.

\begin{figure*}[ht]
\centering 
\includegraphics[width=16cm]{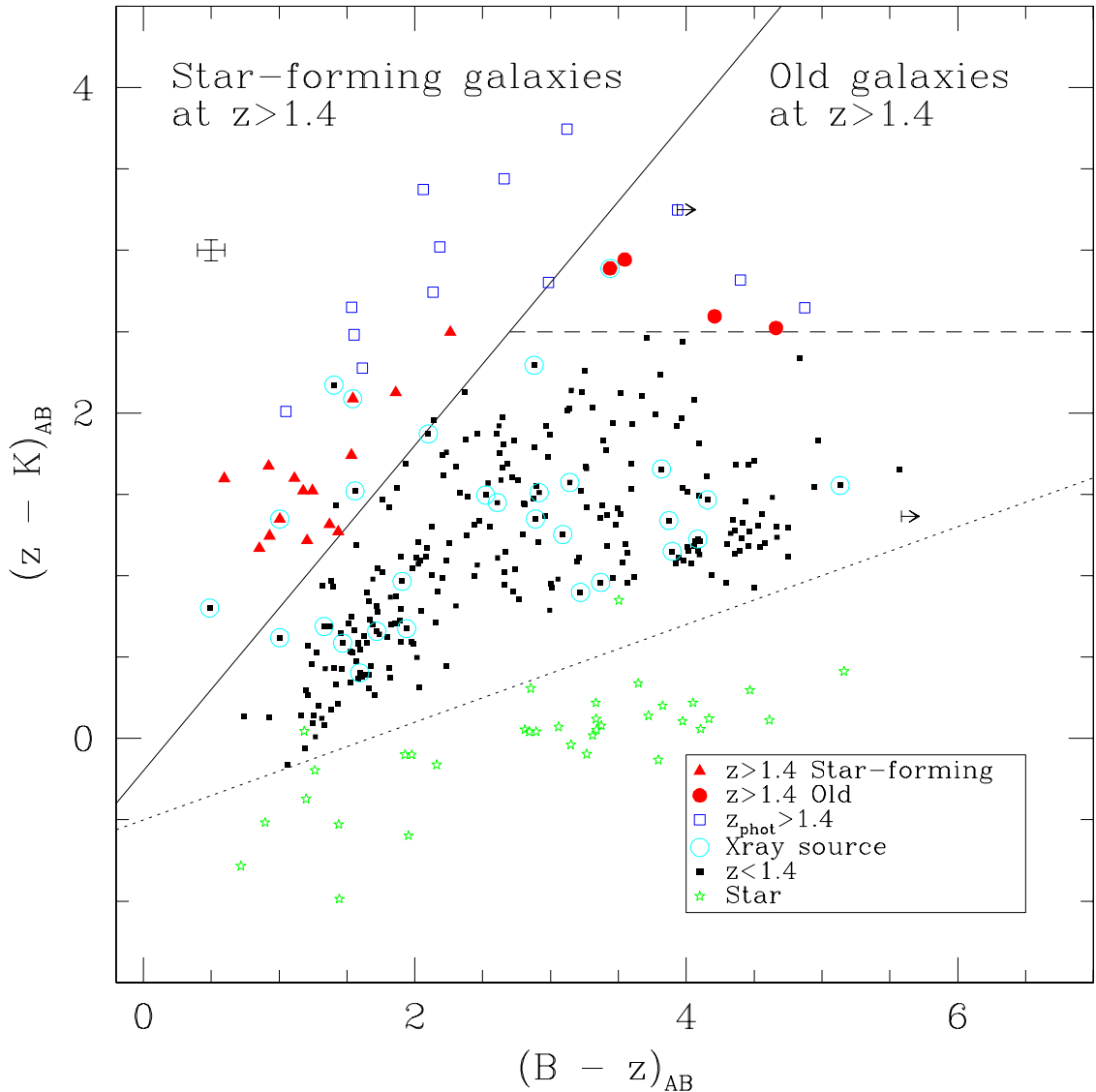}
\caption{
Two color $(z-K)$ vs $(B-z)$ diagram for the galaxies in the
GOODS area of the K20 survey. 
Galaxies at high redshifts are highlighted: solid triangles represent
galaxies at $z>1.4$ with features typical of young star forming systems
(D04);
solid circles are for $z>1.4$ galaxies with old stellar populations 
(Cimatti et al. 2004); empty squares are objects with no measured
spectroscopic redshift and $z_{\rm phot}>1.4$. 
Sources detected in the X-ray catalog of
Giacconi et al. (2002) and/or Alexander et al. (2003) are circled.
Stars show spectroscopically identified galactic objects.
The diagonal solid line defines the region $BzK\equiv (z-K)-(B-z)\geq-0.2$ 
that is efficient to isolate $z>1.4$ star forming galaxies. The
horizontal dashed line further defines the region $z-K>2.5$ that
contains old galaxies at $z>1.4$.
The error bar located in the top-left part of the diagram shows the
median error in the $(z-K)$ and
$(B-z)$ colors of objects at $z>1.4$ (either photometric 
or spectroscopic). The dotted diagonal defines the region occupied
by stars.
The four objects with $z_{\rm phot}<1.4$ are not highlighted and occupy the
same region of $z_{\rm spec}<1.4$ objects. 
}
\label{fig:BzK}
\end{figure*}

In addition to optical-IR data, the publicly available deep 1 Msec Chandra 
X-ray observations of the area (Giacconi et al. 2002), and deep VLA 
radio maps (Kellermann et al. 2004), are used. Some details
on the properties of the radio data and data analysis methods were
summarized in Cimatti et al. (2003).

\subsection{The K20/Q0055 Field}

The K20 spectroscopic dataset from the 19 arcmin$^2$ area centered on
the QSO 0055-269 at $z=3.656$ (Q0055 hereafter) was also used as a 
valuable additional sample (Cimatti et al. 2002b).
The spectroscopic completeness in the area is lower (89\%).
Of the 198 objects, 176 have a spectroscopic redshift identification at the
moment, including 167 extragalactic objects and 9 stars. 
Of these, 13 lie at  $z>1.4$ (8\% of the galaxies) and 12 have
$z_{\rm phot}>1.4$, for a total of 15\% expected at $z>1.4$. 
The imaging (described in full detail in Cimatti et al. 2002b)
has worse seeing (generally $\sim1''$) and
is shallower than in the K20/GOODS area, as the data were mainly 
obtained at the ESO NTT 3.5m telescope (with SUSI2 and SOFI instruments),
except for the Gunn $z$-band obtained with the VLT+FORS1 which has a 
similar depth and seeing to the one of the K20/GOODS area (although shallower than 
the ACS $z$-band imaging). For the above reasons, the photometry
is less accurate for this field. In particular for the $BzK$
bands it turns out that in general the galaxies still have very good
photometry in the $K$- and $z$-bands but the reddest galaxies have quite
poor $B$-band photometry in the Q0055 field, as the 5$\sigma$ limits in 
the photometric apertures are $\sim26.0$ and $\sim27.6$ AB magnitudes for
the K20/Q0055 and K20/GOODS regions respectively.
X-ray, radio and HST 
data are not available for this field.

\section{Near-IR color selection and classification
of $1.4\simlt z\simlt2.5$ galaxies}
\label{sect:BzK}

\subsection{The $BzK$ Criterion}
Fig.~\ref{fig:BzK} shows the $B-z$ vs. $z-K$ colors of the 311 galaxies and
the 36 stars in the K20/GOODS sample. The classification of galaxies at
$z>1.4$ as star-forming objects relies on [OII]$\lambda3727$ emission 
($1.4<z\simlt1.7$), while those at $z>1.7$ have UV spectra showing the 
typical features of star-forming galaxies including e.g. the CIV
absorption system at 1550 \AA\ (D04, De Mello et al. 2004).
It is found that $z>1.4$ star-forming galaxies occupy a narrow 
range and well defined region in this plane, 
well separated by lower redshift galaxies, with the bluest
$B-z$ color at fixed $z-K$. 
By  defining:
\beq
BzK\equiv (z-K)_{AB}-(B-z)_{AB},
\label{eq:BzK}
\eeq
it follows that $z>1.4$ star-forming galaxies are all  selected by the 
criterion:

\beq
BzK\geq -0.2,
\label{eq:cond1}
\eeq
i.e., to the left of the solid line in Fig.~\ref{fig:BzK}. 
In Fig.~\ref{fig:BzK} are also marked the spectroscopically 
confirmed passive systems at $z>1.4$. The classification of these 
old galaxies relies on the detection of significant continuum 
breaks and absorption features in the rest-frame 2500--3000 \AA\ region 
(Cimatti et al. 2004). Being the reddest objects in both $B-z$ and 
$z-K$ colors, old stellar systems at $z>1.4$ can also be readily isolated 
in a BzK diagram using:

\beq
BzK<-0.2\ \ \ \bigcap\ \ \ (z-K)_{AB}>2.5. 
\label{eq:ES0}
\eeq

\begin{figure}[ht]
\centering 
\includegraphics[width=8.8cm]{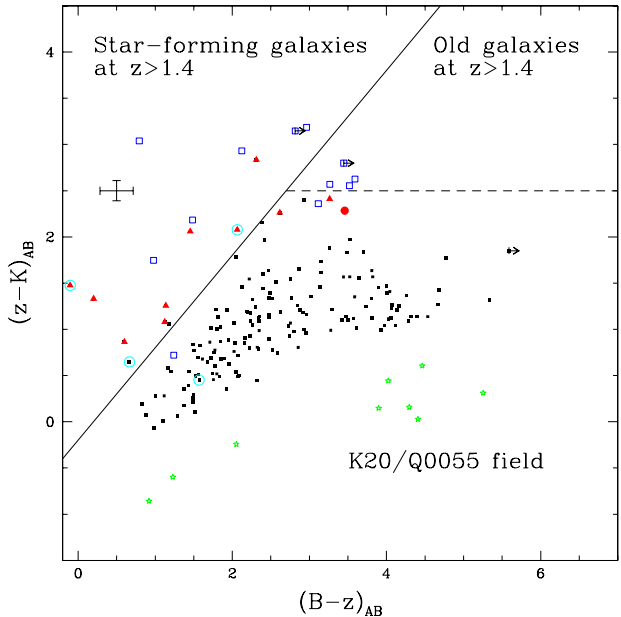}
\caption{The $BzK$ diagram for the Q0055 field of the K20 survey.
Symbols are as in Fig.~\ref{fig:BzK}, except that here
circled symbols represent objects with a Type 1 AGN classification
based on the optical spectra (X-ray data are not available for the
Q0055 region).  }
\label{fig:Q0055_BzK}
\end{figure}

\begin{figure}[ht]
\centering
\includegraphics[width=8.8cm]{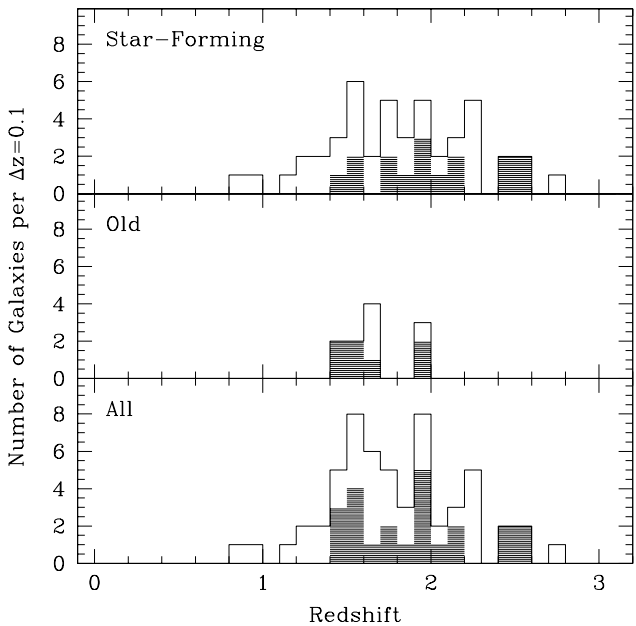}
\caption{
The redshift histogram of the 57 K20 galaxies selected with the criteria 
defined in Section \ref{sect:BzK}. The shaded areas are for
objects with photometric redshift only. The bottom panel shows the redshifts
for all
galaxies, center panel for the old objects and top panel for the
star-forming ones. The contamination of galaxies at $z<1.4$
is only 12\% of the sample and often consists of $z\sim1$ X-ray luminous
galaxies, likely AGN.
}
\label{fig:zhisto}
\end{figure}

\begin{figure}
\centering 
\includegraphics[width=8.8cm]{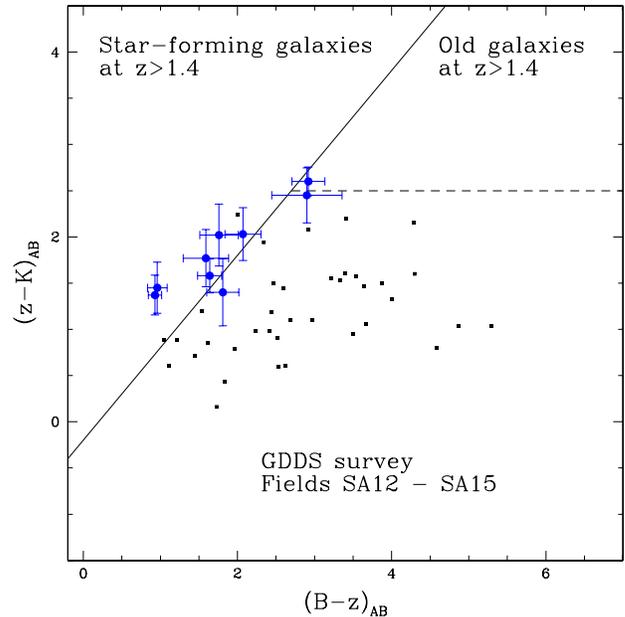}
\caption{The $BzK$ diagram for galaxies in the GDDS survey
with $K<20.6$ (Abraham et al. 2004). Large symbols are for galaxies 
with spectroscopic redshift $1.4<z<2.2$, small squares for $z<1.4$ galaxies.
}
\label{fig:GDDS}
\end{figure}

\begin{figure}[ht]
\centering 
\includegraphics[width=8cm]{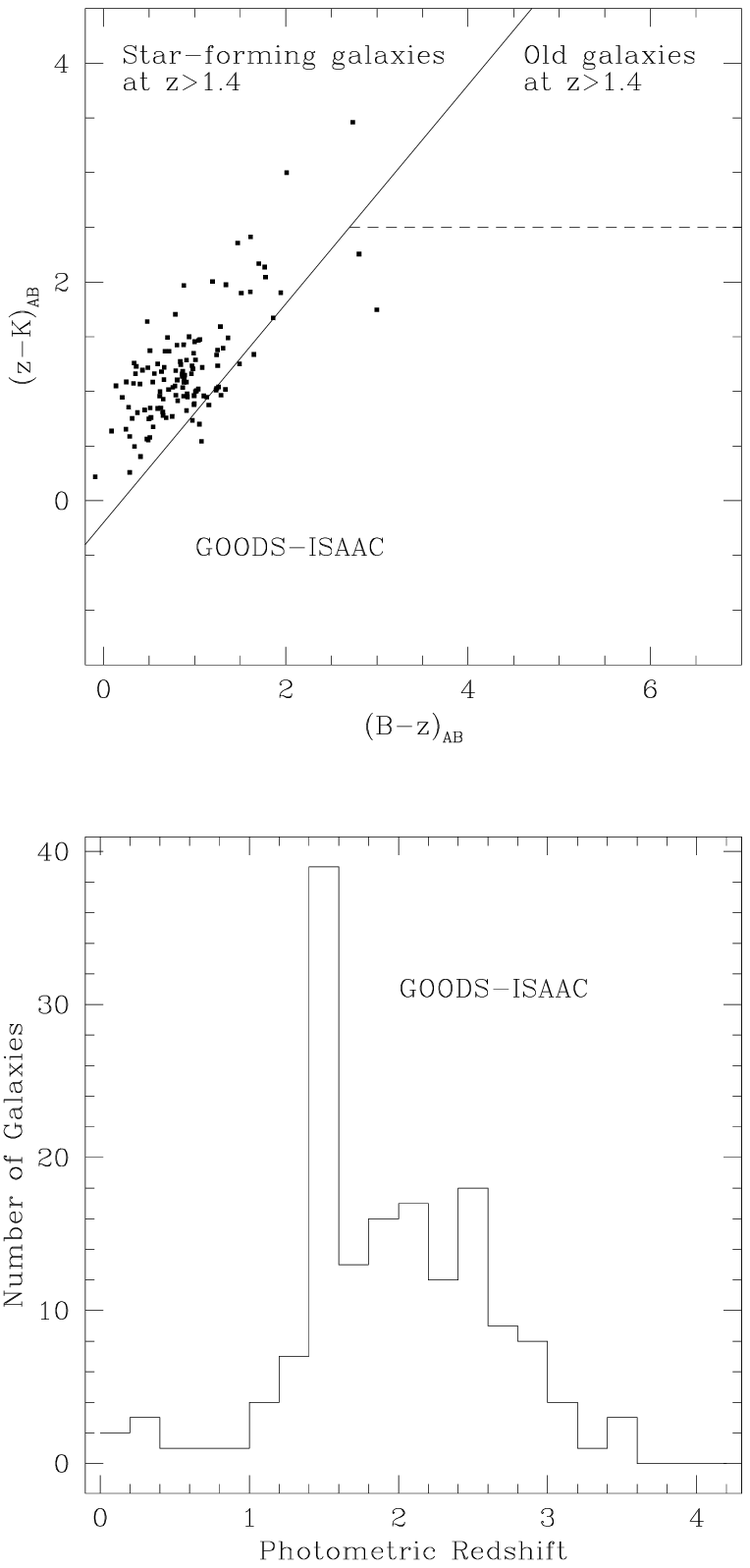}
\caption{Top: the $BzK$ diagram for galaxies with $20<K<22$ 
in the GOODS-ISAAC region selected with $1.4<z_{\rm phot}<2.5$.
Bottom: the photometric redshift distribution of galaxies with
$20<K<22$ in the GOODS-ISAAC region selected with the $BzK$ criteria.
It is not clear whether the narrow spike at $z_{\rm phot}\simeq 1.5$
is real or just an artifact of photometric redshifts.  }
\label{fig:IS2}
\end{figure}

All objects with $z_{\rm phot}>1.4$  are also
selected by the above criteria, as evident from  Fig.~\ref{fig:BzK}. 
Thus, the overall $BzK$-selected sample includes 25 star-forming galaxies at $z>1.4$
having $BzK\geq-0.2$ (15 $z_{spec}$ and 10 $z_{\rm phot}$) and 7 old galaxies at $z>1.4$
having $BzK<-0.2$ and $z-K>2.5$ (4 $z_{spec}$ and 3 $z_{\rm phot}$).
The above criteria are therefore quite efficient in singling out 
$z>1.4$ galaxies, as the lower redshift {\it interlopers}
are only 13\% of the resulting samples, i.e. 5 objects (including 3 Chandra
sources at $0.8<z<1.2$ and 2 star-forming galaxies at 
$1.2<z<1.4$). 
It is not unexpected that X-ray luminous objects, i.e.
AGN, may contaminate these samples as a similar selection technique to
the one devised here was proposed to identify luminous QSOs (Sharp
et al. 2002). A QSO at $z=2.8$ (not highlighted in Fig.~\ref{fig:BzK}) 
is instead not selected by the method. Stars have colors that are clearly
separated from the regions occupied by galaxies (and in particular by those at
$z>1.4$), and can be efficiently isolated with the criterion: 
$(z-K)<0.3(B-z)-0.5$ (dotted diagonal line in Fig.~\ref{fig:BzK}).

The Q0055 dataset was used as an independent verification for the 
validity of the $BzK$ selection. 
Fig.~\ref{fig:Q0055_BzK} shows the resulting $B-z$ versus $z-K$ diagram 
analogue to Fig.~\ref{fig:BzK}. Also in this field
18/23 of the
galaxies with either spectroscopic or photometric redshift
$1.4<z\simlt2.5$ are selected by the method. A few $z>1.4$ objects 
remain marginally out of the $BzK$ selection regions. Most of these 
have either $1.4<z<1.5$ or very poor $B$-band photometry and all
are consistent with lying in the selection regions within 1--$1.5\sigma$.
Comparison of Fig.~\ref{fig:BzK} and \ref{fig:Q0055_BzK} clearly shows
that the objects in the $BzK>-0.2$ region are more scattered out in
the K20/Q0055 field than in the K20/GOODS area because of the worse
quality of the photometry. 
Two objects, including a galaxy and an AGN at $z>3$ (not highlighted in
Fig.~\ref{fig:Q0055_BzK}), are 
not selected by the criteria, that appear to have its main efficiency
at $1.4<z\simlt2.5$, as justified from modeling in
Section \ref{sec:modeling}.
The contamination from low-redshift galaxies is also here quite reduced.
There are 2 objects having $BzK>-0.2$ that lie at $z<1.4$ and result to be 
a $z=1.367$ star-forming galaxy and an AGN at $z=1.119$. 
The criteria appear quite successful on the Q0055 dataset as well, 
once accounting also for the overall lower quality of the  dataset,
as discussed above. 

These criteria thus allow a very efficient and highly complete
selection of the 55 galaxies with either
spectroscopic or photometric redshift $1.4<z<2.5$ in the K20 survey.
The 57 galaxies selected with the $BzK$ criteria to $K<20$ correspond to
a surface density of about $1.1\pm0.15$ arcmin$^2$ and have a redshift
distribution mainly spread over $1.4<z<2.5$ (see Fig.~\ref{fig:zhisto}). 

\subsection{$BzK$-selected Galaxies in the GDDS and GOODS Fields}

The $BzK$ selection was applied to other available samples in order to
further verify its validity and test it at magnitudes fainter than $K\sim20$.

The Gemini Deep Deep Survey (GDDS; Abraham et al. 2004) performed spectroscopy
of galaxies selected to $K<20.6$ (Vega). They spectroscopically observed 
a fraction of
their $K$-selected sample favoring the objects with the redder colors. 
Two of the GDDS fields (SA12 and SA15) have $BzK$ photometry.
After converting the GDDS $B$ and $K$ band photometry 
from Vega to AB scale, Fig.~\ref{fig:GDDS} shows that 7/9 GDDS galaxies 
with $z>1.4$ can be selected with the $BzK$ technique, with only two 
contaminants
from $z<1.4$. Two galaxies with $z>1.4$ are just outside the
$BzK$ selection regions. The photometric errors in the $BzK$
magnitudes in such a catalog are on average significantly larger 
than for the K20/GOODS galaxies with $z>1.4$. Within the errors, also the
two outliers are consistent with the $BzK$ criterion.

As a further check, the galaxies within the deep ISAAC imaging of GOODS
were considered at depths $K>20$ and up to $K=22$, 
i.e. two magnitudes fainter than reached by the K20
survey, for the same 32 arcmin$^2$ region covered by the K20 survey.  
Only the objects with well determined SEDs were included,
requiring errors smaller than 0.3, 0.15, 0.15 mags for the $B$-, $z$-, and
$K$-bands, respectively.  This ensures reasonably reliable photometric redshift
determinations, that were obtained using {\em hyperz} in a similar way
as for the brighter K20 galaxies and using the same $UBVRIzJHK$ imaging
datasets. The $BzK$ colors of the 125
galaxies selected to have $1.4<z_{\rm phot}<2.5$ 
are shown in
Fig.~\ref{fig:IS2}, where they indeed concentrate in the region with
$BzK>-0.2$. About $10$\% of them are just marginally
outside the $BzK$ selection region.
No red, passively evolving galaxies and very few red $BzK>-0.2$
galaxies are identified in the sample,
most likely because of the adopted, stringent criterion on 
the photometric errors. Fig.~\ref{fig:IS2} also shows
that the photometric redshift distribution of the 159 galaxies
selected with the $BzK$ criteria 
(for a lower limit to their sky density of $\simgt5$ arcmin$^{-2}$ at $K=22$) 
is indeed centered at $z\sim2$. Only
about 10\% of the galaxies are at $z_{\rm phot}<1.4$, while 15\% of them
has $z_{\rm phot}>2.5$.  The galaxies in the deep samples of
Fig.~\ref{fig:IS2} have a median of
$K(Vega)=21.2$. 

These two checks are satisfactory and support the idea that
the method is valid also for samples selected at magnitudes somewhat fainter
than $K=20$. Such a validity should be further tested with future
surveys.

\begin{figure*}[ht]
\centering 
\includegraphics[width=16cm]{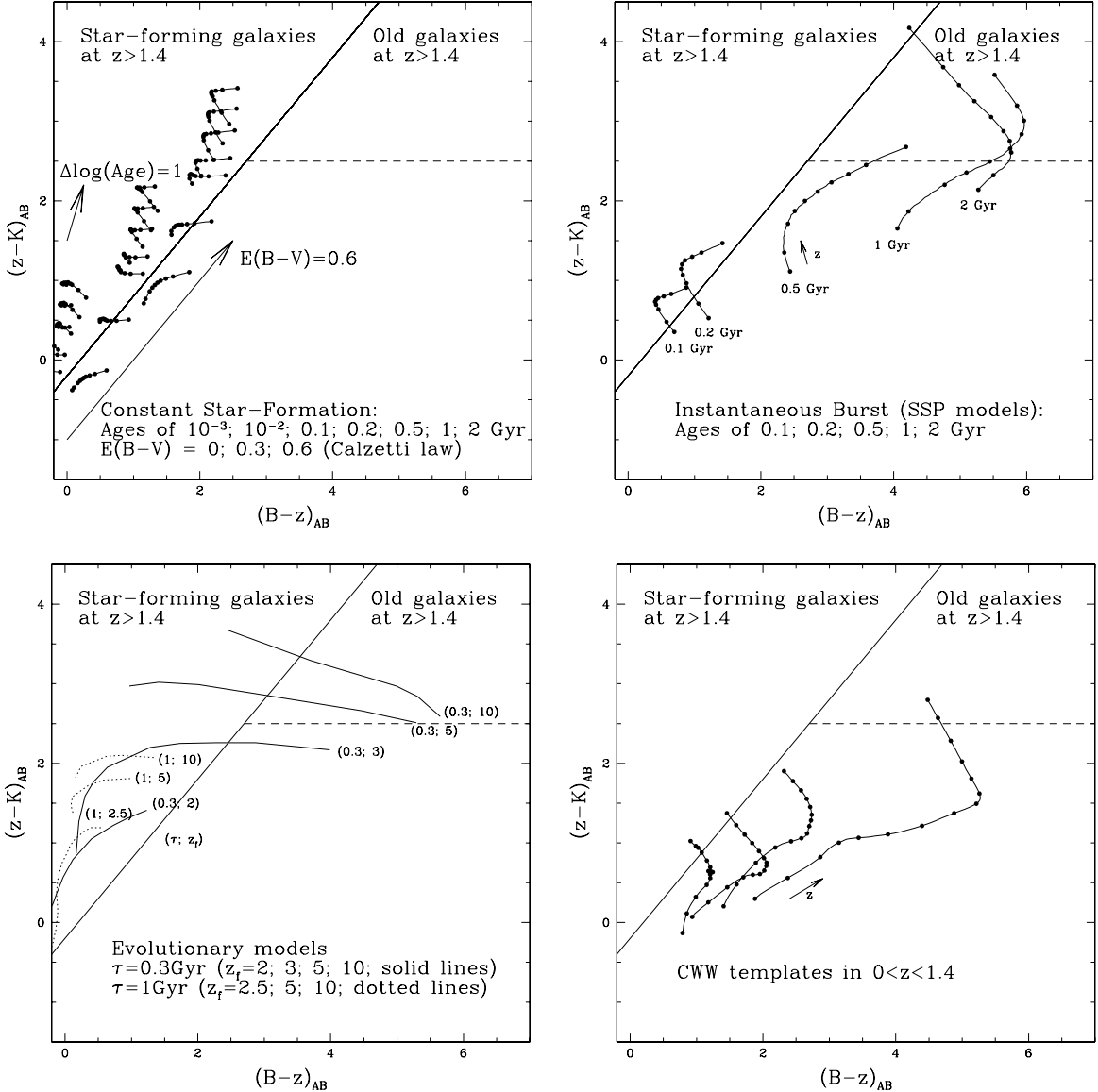}
\caption{The evolutionary tracks in the $BzK$ diagram from  theoretical 
models.The top-left panel shows continuous star formation model tracks
for ages from 1 Myr to 2 Gyr and for E(B-V)=0, 0.3, 0.6.  Top right
panel has simple stellar population models for ages from 0.1 to 2 Gyr
and no reddening. The bottom-left panel shows evolutionary models with
various formation redshifts and SFR timescales and no reddening.  At
decreasing redshifts the tracks generally turn from bottom-left to
top-right. The bottom-right panel show colors for the local templates of
various galaxy types from Coleman et al. (1980, CWW).  All models are
plotted for the range $1.4<z<2.5$, except in the bottom-right panel where
plots are for $0<z<1.4$.  The limits and color ranges of all four
panels reproduce those of Fig.~\ref{fig:BzK}, for direct reference.  }
\label{fig:4p}
\end{figure*}

\begin{figure}
\centering 
\includegraphics[width=8.8cm]{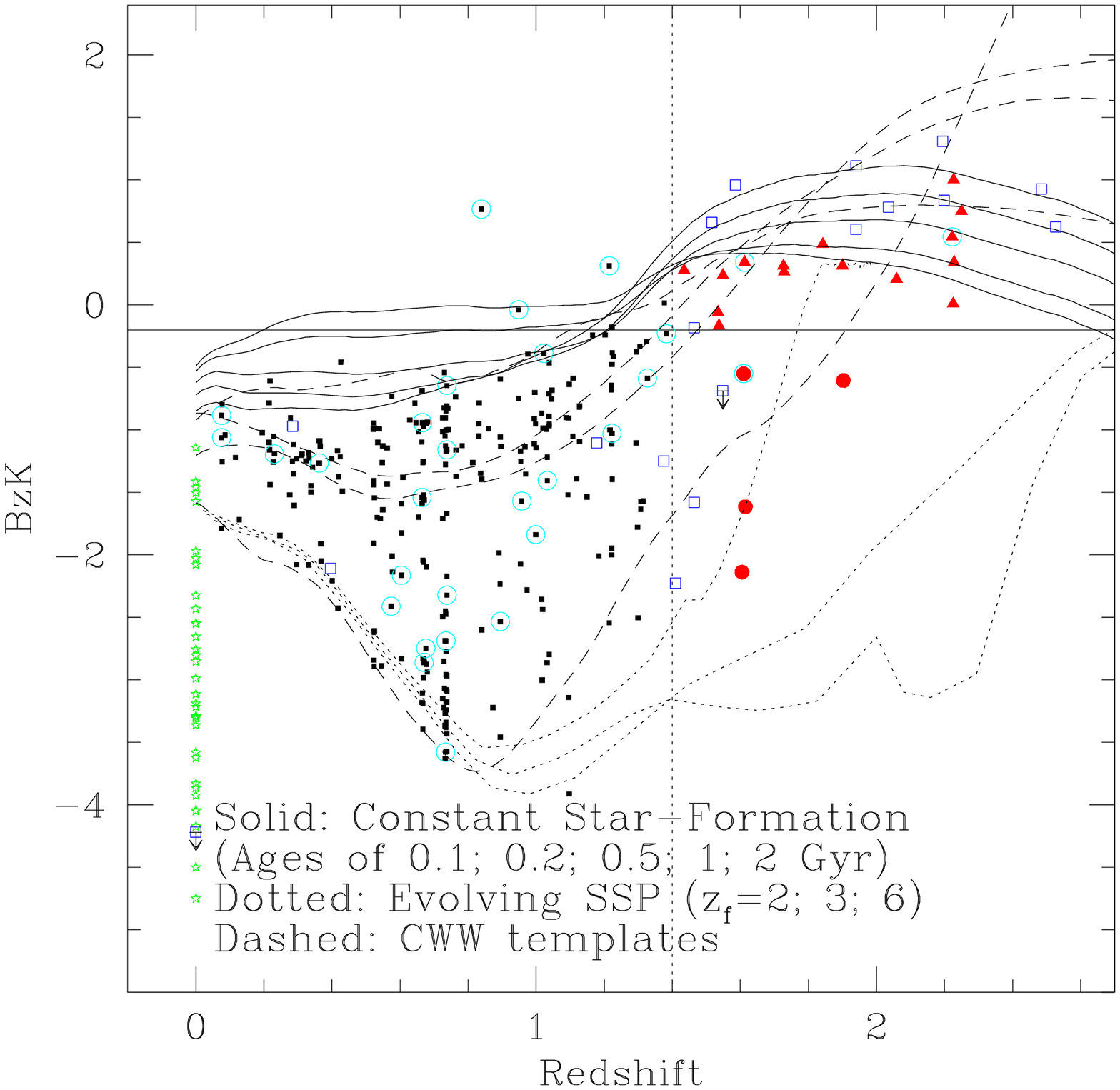}
\caption{$BzK$ versus redshift for galaxies in the K20/GOODS area. 
Symbols are as
in Fig.~\ref{fig:BzK}, except that here all objects with photometric 
redshift only are shown as empty squares.
Model tracks are also over-plotted showing
the expected $BzK$ color as a function or redshift. Constant
star-formation rate models (solid lines) 
are shown for ages between 0.1 and 2 Gyr and
reddening $E(B-V)=0.3$ (but note that the $BzK$ color is nearly reddening
independent at $z>1$). Also shown is color evolution for  evolving stellar
populations formed in an instantaneous burst at redshifts $z=2$, 3 and 6
(dotted lines) and the variation of  color with redshift for the 
(non evolving) templates of E, Sbc, Scd, and Im local 
galaxies from Coleman et al. (1980, CWW).
}
\label{fig:BzKvsz}
\end{figure}

\section{Stellar population modeling}
\label{sec:modeling}

Bruzual \& Charlot (2003) stellar population synthesis 
models were used to further elucidate the physical meaning and the validity 
of these, phenomenologically established, $BzK$  criteria.

\subsection{Galaxy Models in  the $BzK$ Diagram}

All the four panels of Fig.~\ref{fig:4p} reproduce the $B-z$ and
$z-K$ range of Fig.~\ref{fig:BzK} for different set of models.

The top-left panel
shows the location of constant star-formation (CSF)
models, computed for ages from $10^{-3}$ to 2 Gyr, $1.4<z<2.5$, various
levels of reddening (using the Calzetti et al. 2000 extinction law),
and solar metallicity.  The figure confirms that galaxies in such a
redshift range with ongoing star formation are indeed expected to lie
in the $BzK>-0.2$ region of the diagram. 
The duration of the
star-formation (age) has little influence on the $B-z$ color, while
$z-K$ increases with age, an effect due to the development of strong
Balmer/4000 \AA\ breaks falling beyond the $z$-band for $z\approx2$.  
Very young bursts with age
less than 10 Myr (and no underlying older stellar populations) would
be located just below the threshold. Galaxies with similar properties
are not present in the K20 database ($K<20$) and are at most only a
minority in the GOODS photometric redshift sample to $K=22$
(Fig.~\ref{fig:IS2}). In order to include also such objects one should 
formally require $BzK\simgt-0.8$. In the K20 sample, this would increase 
significantly the contamination by $z<1.4$ galaxies.

The reddening vector is virtually parallel to the $BzK=-0.2$ limiting line,
implying that dust affects the $B-z$ and $z-K$ colors by {\em the same
amount} for $1.4<z<2.5$ galaxies. This appears to be the reason why the
$BzK>-0.2$ criterion is successful at identifying $z>1.4$ star forming
galaxies, virtually regardless of their dust reddening.  
It can be noticed that
for $E(B-V)=0$ the tracks lie near the edge of the diagram, where
no galaxies are found in the K20 sample (see Fig. 3), suggesting that 
purely unreddened, star-forming
galaxies
are rare at least in a $K$-selected sample at the relatively bright
$K<20$ limit. Instead, in the $K<22$ sample from GOODS
(Fig.~\ref{fig:IS2}), some fainter objects start to occupy also the
region with $0<(z-K)_{AB}<1$.

The top-right panel of Fig.~\ref{fig:4p}
shows the location of simple stellar population (SSP)
models in the $BzK$ diagram, for ages of 0.1 to 2 Gyr, no
reddening, and solar metallicity.  For young ages ($\simlt0.2$) Gyr
the tracks are marginally redder but similar to those of star-forming
galaxies. As ages grow to $\simgt 1$ Gyr the tracks occupy the region
where passive $z>1.4$ galaxies are detected, as expected.  There is an
intermediate age regime at $\sim0.5$ Gyr for SSP models in which such
objects would be missed by both criteria of Section~\ref{sect:BzK} for
$z\approx 2$. 
SSP models with young ages
(i.e. comparable to the duration of major star formation events in
real galaxies) and no reddening might be an unrealistic
schematization, as real young galaxies are likely to be to some
extent still star-forming and dust-reddened.

As a more reasonable rendition  and in order to explore the redshift
and aging effects, some evolutionary models were computed with star
formation histories more extended in time,
as described in Daddi et al. (2000b). 
Left-bottom panel of Fig.~\ref{fig:4p} 
shows the $BzK$ colors in $1.4<z<2.5$ for galaxies
with various formation redshifts and exponentially declining SFRs
($\tau=0.3$ and 1 Gyr), with no reddening
and solar metallicity.
For high formation redshifts ($z_f>5$) the implied color evolution
at $z>1.4$ is such that objects move
directly from the star-forming galaxy region ($BzK>-0.2$) to the 
passive galaxy region ($BzK<-0.2$ and $z-K>2.5$) without crossing
the bluer regions populated by $z<1.4$ objects. 

The last panel of Fig.~\ref{fig:4p} finally shows that the color of
normal galaxies at $0<z<1.4$, computed using the Coleman et al (1980)
templates of E-Sbc-Scd-Irregular galaxies, are indeed
expected to fall outside the range defined for $z>1.4$ and bracket
quite well the range of colors observed for K20 galaxies at $z<1.4$ 
(see Fig.~\ref{fig:BzK}). This additional test strengthens the validity
of the $BzK$ selection to isolate galaxies at $z>1.4$. 

\subsection{Modeling $BzK$ versus Redshift}

The key quantity in the selection and classification of $z>1.4$ galaxies 
is the $BzK$ term defined in Eq.~\ref{eq:BzK}. 
Fig.~\ref{fig:BzKvsz} shows the $BzK$ evolution as a function 
of redshift for the CSF models described above.
Objects with $BzK\simgt0$ start to appear in
significant numbers only beyond $z>1.4$.
The CSF models enter the region $BzK>-0.2$ at
$z\simgt1.2$, although models with ages of 1--2 Gyr can
marginally fulfill the $BzK>-0.2$ condition even at much lower redshifts. 
CSF models evolve out of the $BzK>-0.2$ region at
$z>2.6$--3.2, depending on age, because the Ly$\alpha$ forest starts entering
the $B$-band at those redshifts, thus  producing
a reddening of the $B-z$ colors.
Also shown in the figure are $BzK$ colors expected for passively evolving
(SSP) galaxies formed at $z=2$, 3 and 6,
and templates of local galaxies (Coleman et al. 1980). 
Again, one notices that in general passively evolving
galaxies are contaminating $BzK>-0.2$ samples only for very young
ages close to the formation redshift. 
The Coleman et al. (1980) templates
bracket the $BzK$ color range observed for $0<z<1$ galaxies,
as well as that of higher redshift $z\sim2$ galaxies (although
they have too old stellar populations to be truly representative of 
$z\sim2$ galaxies).

\subsection{The Effects of Metallicity and Extinction Laws}

We also explored  the effects of using alternative choices for the 
metallicity and the extinction law.
Models with 
metallicity significantly below solar seem inappropriate even for $z\sim2$ 
star-forming
systems with $K<20$, as these objects show deep photospheric absorption spectra
indicative of solar or higher metallicity (De Mello et al. 2004). Old
systems are consistent with being fully assembled spheroids, that are known to
have nearly solar or higher metallicity today.  We checked that
using above-solar metallicities  the CSF galaxy tracks are
basically unchanged, while the tracks of passive galaxies change
according to the well known age/metallicity degeneracy.
We also investigated 
the effect of using extinction laws other than  that of  Calzetti et al.
(2000).  Using the
extinction law proposed by Silva et al. (1998) yields results fully
consistent with those obtained above with the Calzetti et al. 
law. The SMC extinction curve produces a higher reddening to the $B-z$
color than to the $z-K$ color, so that for very high reddening
($E(B-V)\simgt0.6$) the model tracks would enter the $BzK<-0.2$ region
of passive galaxies. However, with the SMC extinction law the colors 
of the reddest galaxies with $BzK>-0.2$ are difficult to reproduce.

\begin{figure*}
\centering 
\includegraphics[width=8cm,angle=-90]{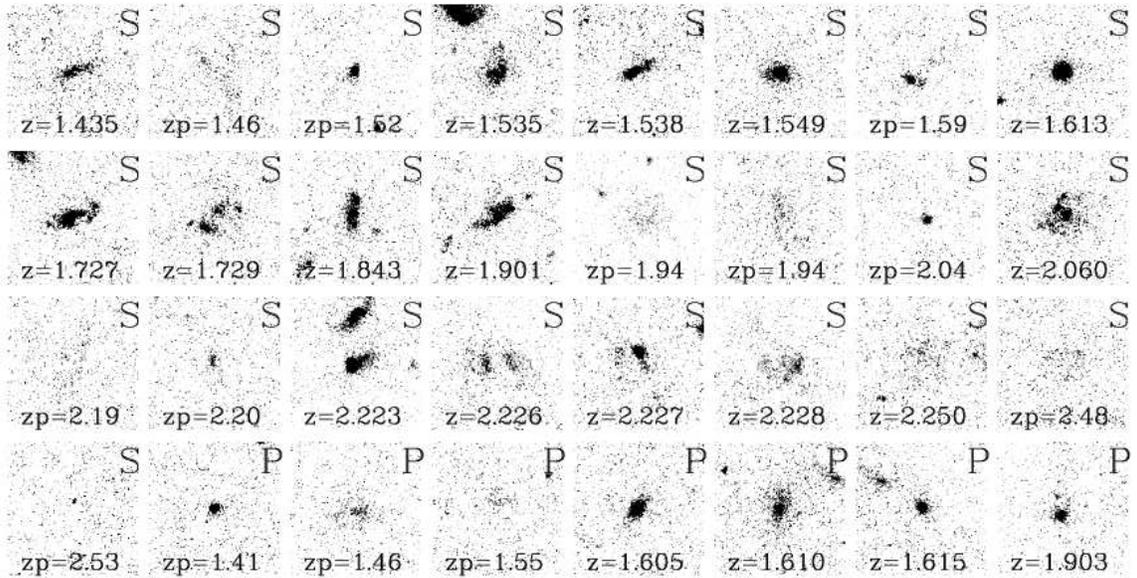}
\caption{
ACS z-band (F850LP) snapshot images ($5^{\prime\prime}\times
5^{\prime\prime}$) of the 32 K20/GOODS galaxies at $z>1.4$ that can be selected
with the $BzK$ criteria. The galaxies are divided between candidates 
star-forming ("S", defined as those with $BzK\geq-0.2$) and passive ("P", 
defined as those with $BzK<-0.2$ and $z-K>2.5$), according to the diagnostic
discussed is Section~\ref{sect:BzK}. Galaxies in each category are sorted by
increasing redshift ("zp" means that the redshift is photometric).
A subsample of these images had been shown in D04 and Cimatti et al
(2004).
}
\label{fig:morphoBzK}
\end{figure*}

\subsection{The Nature of the Reddest Galaxies with $BzK>-0.2$}

Most of the K20 galaxies with no spectroscopic redshift available 
and $1.4<z_{\rm phot}<2.5$ have very red $(z-K)_{AB}>2.5$ colors and
$BzK>-0.2$ (Fig.~\ref{fig:BzK}), qualifying thus as star-forming galaxies based
on the proposed classification criteria. The top-left panel of
figure~\ref{fig:4p} confirms that such objects
are fully consistent with being heavily reddened, star-forming galaxies.  
Their full multicolor SEDs cannot generally 
be fitted by models for old/passive galaxies with no star-formation and
reddening, implying that 
some amount of young-hot stars is required for them to show
the relatively high $B$-band fluxes and blue $B-z$ colors.  
This is quite reasonable,
as they appear to follow the trend of increasing reddening for the
spectroscopically established star-forming galaxies at $z>1.4$.
Nevertheless, some of the objects in that region may actually be 
post-starburst galaxies,
having passed their strongest episode of star-formation, in
which case part of their red colors could be due to an aged burst of
star-formation (see bottom-left panel of Fig.~\ref{fig:4p}).
Additional
evidence that, typically, these are indeed actively star-forming galaxies 
will be derived from their average X-ray and radio properties in
Section ~\ref{sec:Xr}.

\section{HST/ACS Morphology of the $z>1.4$ Galaxies}
\label{sect:HST}

HST imaging provides a fundamental complement to investigate the 
nature of the $BzK$ galaxies and to elucidate their evolutionary status. 
In Fig.~\ref{fig:morphoBzK} ACS z-band imaging of the $z>1.4$ galaxies
in the K20/GOODS sample are presented. The $z$-band is centered at
rest-frame wavelengths from
2500 \AA\ to 3700 \AA\ for the objects in $1.4<z<2.5$.
Objects with $BzK\geq-0.2$ appear generally irregular/merging-like and
have very large sizes, with an average half-light radius of
about 6 kpc at $z\sim2$ (D04). Objects with
$BzK<-0.2$ and $z-K>2.5$ have instead generally a compact and
regular morphology. With only a few exceptions, there is a very
good agreement of the {\it early-type/late-type} morphological
appearance with both the $BzK$ color classification and with the
spectroscopic classification as passive or star-forming galaxies.
This supports the evidence that the $BzK$ criteria allow to
efficiently isolate high redshift galaxies in a $K$-selected sample
and to distinguish passive and star-forming objects. 

\section{Star-formation Rates}
\label{sec:SFR}

The critical
question is then to investigate the level of star formation activity
present in the $BzK$ (star-forming) galaxies. This is done in this Section
where we measure the SFRs of galaxies at $1.4<z<2.5$ in the K20/GOODS region
(hence with $K<20$), 
selected and classified as star-forming 
with the criterion $BzK>-0.2$, and we explore whether the $BzK$ 
photometry alone can still provide an estimate of the SFRs of these galaxies. 

\subsection{SFRs from the Rest-Frame UV-Continuum Luminosity}
\label{sect:ebv}

Estimating the SFR of a galaxy from its multicolor optical photometry
is generally based upon relations between the intrinsic UV continuum
luminosity
and SFR (e.g. Madau et al. 1998) and estimating the extinction by dust,
necessary to derive the intrinsic UV continuum luminosity from the observed 
one. Both steps are in general quite uncertain and rely on assumptions about 
star-formation history, dust reddening law, metallicity (as well as on the
IMF, fixed to Salpeter between 0.1 and 100 $M_\odot$ in this work, but in a way that is generally easy 
to factor-out so that the results can be easily scaled to other choices
of IMF).

Neglecting any reddening 
correction, the SFRs are of order of $\approx10$--40~$M_\odot$yr$^{-1}$ 
for the spectroscopically confirmed galaxies (D04), and even smaller for the 
reddest ones with only photometric redshifts. As well known, such
estimates are severely affected by dust extinction. In this section,
we attempt to infer the level of dust reddening from the photometric
properties,
limiting this analysis to the case of CSF models with solar metallicity 
and a Calzetti et al. (2000) extinction law (as already done in D04). 
These assumptions appear reasonably justified for our galaxies, as 
discussed in Section ~\ref{sec:modeling}, and allow a comparison with a 
broad variety of literature work based on the same assumptions.

With the above assumptions and following the approach of D04, best-fitting 
SFRs and $E(B-V)$ 
have been derived for the 24 purely star-forming galaxies with 
$z>1.4$ in the K20/GOODS area, from their full observed SED 
from $U$ to $K$ (an object with AGN signatures in the spectrum and 
high X-ray to optical luminosity ratio has been excluded from
the SFR analysis). The derived SFRs  are typically in
100--600~$M_\odot$yr$^{-1}$ and the reddening range is $0.2\simlt
E(B-V)\simlt1$. 

\begin{figure}[ht]
\centering 
\includegraphics[width=8.8cm]{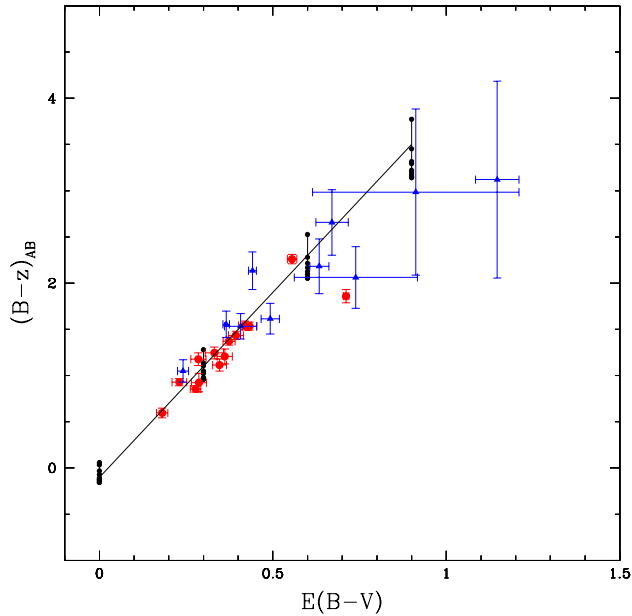}
\caption{
The $B-z$ color is plotted vs. the reddening $E(B-V)$ for the
$BzK$-selected star-forming galaxies in the K20/GOODS region. 
The best fitting $E(B-V)$ values
from the full SED analysis are shown for individual objects. Circles
with error bars: objects with spectroscopic redshifts; triangles:
objects with only photometric redshifts.  Also shown is the $B-z$
color vs.  $E(B-V)$ for constant star-formation rate, 500 Myr old
models and various redshifts within the range $1.4<z<2.5$ (filled
circles connected by vertical lines). The 500 Myr age is the typical SED
best-fitting age for K20 $z\sim2$ star-forming galaxies (D04).  
The diagonal
line shows the relation defined in Eq.~\ref{eq:redde}.  }
\label{fig:redde}
\end{figure}

Then, we have explored if, within the same assumptions, the
$BzK$ photometry alone could allow an estimate of the
SFR content of the 24 galaxies equivalent to the one derived from SED fitting.
At $z\sim 2$ the $B$-band samples quite well the rest-frame
ultraviolet at $\sim 1500$ \AA\ (actually, the rest-frame 1250--1800
\AA\ range for $1.4<z<2.5$), and the UV luminosity at 1500 \AA\ of a
star-forming galaxy is a calibrated measure of the ongoing
SFR (e.g. Madau et al. 1998). Fig.~\ref{fig:redde} 
suggests that $E(B-V)$ estimated from SED fitting (and with the
knowledge of the galaxy redshift) correlates very well with the 
observed $B-z$ color
(see also Fig.~\ref{fig:4p}, top left panel), following the relation:

\beq
E(B-V)=0.25(B-z+0.1)_{AB}
\label{eq:redde}
\eeq
that indeed for the
models and assumptions described above holds as an average over redshift and 
age.
The rms dispersion of the residuals is
only 0.06 in $E(B-V)$ for the objects with measured spectroscopic
redshift (mainly due to a single outlier, with dispersion dropping to
0.026 with such object removed), and the relation still holds quite
well also for objects with photometric redshift only. This tight
relation between $E(B-V)$ and $B-z$ is related to the assumption of
the "grey" and self-similar Calzetti et al. (2000) extinction law
and to the fact that the UV shape of CSF models has little dependence
on age.

The observed flux at 1500 \AA\ rest-frame, 
de-reddened using Eq.~\ref{eq:redde}, can be used to estimate the 
1500 \AA\ luminosity once the redshift is known. 
For the BC03 models this can be converted into a SFR using the relation:

\beq
{\rm SFR} (M_\odot{\rm yr}^{-1}) = L_{1500 \AA} [{\rm erg\ s^{-1} Hz^{-1}}] / (8.85\times10^{27}). 
\label{eq:madau}
\eeq

For objects lacking a spectroscopic (or even photometric) redshift
identification, this conversion can be done only assuming an average
redshift for the sample galaxies.  For individual objects, if they lie
at lower (higher) redshift than the average both the luminosity
distance and 1500 \AA\ luminosity (based on just the observed $B$-band
flux) will be overestimated (underestimated), typically by factors up
to $\sim 2$.  However, this procedure would still allow to derive
a fairly correct ensemble average of the SFR in the survey when a
statistical sample of galaxies is considered together.

With the above recipes we have derived a $BzK$ color estimate of the
SFR of each galaxy in the following way: 1) the observed
$B$-band flux is used as a measure of the 1500 \AA\ UV continuum flux; (2)
$E(B-V)$ is derived from the observed $(B-z)_{AB}$ color as in
Eq.~\ref{eq:redde}; and (3) the average redshift of the sample $<\!  z\!>=1.9$
is used for all objects to derive the SFR following Eq.~\ref{eq:madau}.

For the 14 K20 star-forming galaxies with  $1.4<z_{\rm spec}<2.3$, the 
SED fitting and the simple $BzK$ color estimates described above are 
all in agreement within a factor of 2, given our range of redshifts,
with a rms fluctuation of  only 20\%. 
The total SFR from the 14 galaxies results to be 3600~$M_\odot$yr$^{-1}$ 
from SED fitting and 3400~$M_\odot$yr$^{-1}$ with the $BzK$-based 
estimate, in excellent agreement between them and indicating a quite 
high average SFR$\sim250$~$M_\odot$yr$^{-1}$ for these  galaxies.
The SFRs derived for objects with only photometric 
redshifts are more uncertain, but also in that case the agreement
among the two estimates is reasonable. The SFR estimated for the
10 galaxies with $BzK\geq-0.2$ and $z_{\rm phot}>1.4$ (likely 
$z\sim2$ star-forming galaxies with high reddening)
is $\sim 1700\; M_\odot$yr$^{-1}$ in total, corresponding to
SFR$\sim170$~$M_\odot$yr$^{-1}$ per object. Averaging over the two
samples yields SFR$\sim210$~$M_\odot$yr$^{-1}$ for the
typical objects among this population of $K$-selected starbursts.

These results show that, within the assumptions made, the total SFR
content of $1.4<z<2.5$ galaxies can be estimated from the $BzK$
photometry alone with an accuracy similar to that reachable by fitting
to the whole SEDs with known spectroscopic redshifts. However, we notice
that assuming exponentially declining star-formation models the
amount of reddening and SFR can be significantly reduced.  

In order to derive more stringent clues, X-ray and radio data were 
also used to derive independent estimates of the SFR unaffected by 
dust extinction and to test the above results. 

\subsection{SFR from the X-ray Luminosity}
\label{sec:Xr}

Alternative measures of the SFR  of galaxies
can be obtained from their X-ray and radio
properties, as the X-ray and radio luminosities
of star-forming galaxies (with no major AGN contribution) are 
proportional to the SFR (e.g., Condon et al. 1992; Ranalli et al. 2003; Nandra et al. 2003). The
X-ray and radio properties also offer an additional opportunity (besides
optical spectra) to check for the presence of AGN contamination.

Two of the K20 objects at $z>1.4$ with $BzK>-0.2$ are listed as
detections in the catalog based on the 1 Msec
Chandra Deep Field South observations (Giacconi et al. 2002). 
One of these is the object with AGN line
optical spectrum, and we already mentioned that this was excluded from
the star-forming galaxy sample.  Another galaxy at $z=2.223$ with a
faint soft X-ray detection is present in the sample.  This is also
detected as a faint radio source at 1.4 GHz and it is consistent with being
a vigorous starburst with SFR$\simgt500$~$M_\odot$yr$^{-1}$ (D04).

The X-ray emission in the observed 
soft (0.5--2 keV) and hard (2--10 keV) bands have been measured at the
position of the remaining 23 star-forming objects to check for other
possible detections.
No other individual detection is found above the $3\sigma$ level. 
The stacked X-ray signal from the 23 individually undetected sources
was then obtained to constrain the average X-ray emission of the
$z\sim2$ star-forming galaxies.  In the soft band $\sim96\pm23$ net
counts are recovered, after background subtraction. We performed
Monte Carlo simulations by placing at random positions in the
X-ray image (excluding regions around known sources) and found that 
the chance probability of recovering 
such a strong signal is $1.7\times10^{-5}$. 
The average 4.4 soft counts
per objects are close to the detection limits of the 2 Msec Chandra
observations in the HDF North (Alexander et al. 2003). 
Performing the analysis separately on galaxies with
or without spectroscopic redshift identification, it is found that the
two samples have not statistically different X-ray properties and both
samples are positively detected at the $\sim 3\sigma$ level in the soft
band.  
On the other hand no significant detection is found from the stacked hard band 
data, constraining the
hardness ratio of the population to be HR$<-0.54$ at the 2-sigma
level. This is
consistent with the low hardness ratio expected for starbursts
galaxies. AGN are found generally to have $-0.5<HR<0.5$ (Szokoly et al. 2004). 
The low average
HR for our sources thus disfavors that the detected soft X-ray emission is
due to low-level AGN activity.  A similar conclusion
is supported by the low average X-ray-to-optical flux ratio of ${\rm
log} (f_{0.5-2 keV}/f_R) \sim -1.5$ and by the lack of AGN signatures in
the spectra (see also D04). The X-ray emission
is therefore most likely due to star-formation.  Using $\Gamma=2.1$
appropriate for starbursts (e.g. Brusa et al. 2002), the counts
correspond, for $<\!z\!>=1.9$, to a rest frame 2--10 keV luminosity of
$L_{2-10 keV}=8.6\times10^{41}$ ergs s$^{-1}$, which
translates into an average SFR$\sim170$~$M_\odot$yr$^{-1}$ (Ranalli
et al. 2003; Nandra et al. 2002).  When adding
back to the sample the individually X-ray detected (non AGN) object one 
obtains an average X-ray luminosity corresponding to an average 
SFR$\sim190$~$M_\odot$yr$^{-1}$, in quite good agreement
with the estimate from the reddening corrected UV luminosities
and constant star-formation rate models.  

\subsection{SFR from Radio Luminosity}
\label{sec:Rr}

Deep radio maps at 1.4 GHz and 5 GHz (Kellermann et al. 2004) were
used to measure the radio properties of the $z\sim2$ star-forming
galaxies in our sample.  The radio data reach rms flux densities
of about 8 $\mu$Jy
at both 1.4 GHz and 5 GHz.  Two of the star-forming
$z\sim2$ galaxies are individually detected at 1.4 GHz at better than
the 3$\sigma$ level.  One of the two is the vigorous starburst with
SFR$\simgt500$~$M_\odot$yr$^{-1}$ also detected in the X-ray and
discussed in D04. The other object has a 1.4 GHz flux density of $\sim
25\,\mu$Jy and is therefore a $\sim 3\sigma$ detection.  No individual
object is detected at 5 GHz.

The average flux density of non-detections has been evaluated in a similar
fashion as was done for the EROs by Cimatti et al. (2003), and using
the same dataset. The radio flux densities were measured at the
nominal optical position for each of the galaxies averaging the flux
density in the beam (3.5$''$) over a range of 1$''$ radius in order to correct 
for possible residual coordinate mismatch. 
An average signal is measured of $7.4\pm1.8\mu$Jy at 1.4
GHz and $1.5\pm1.8\mu$Jy at 5 GHz, for the 22 $z>1.4$ $K$-selected
star-forming galaxies that are individually undetected both in radio
and X-ray.  The above flux densities do not strongly constrain the radio
continuum slope $\alpha$, but are consistent with the value of
$-\alpha\sim0.6$--0.8 typical of starburst galaxies (Condon et
al. 1992).  For consistency with the work at $z=2$ of Reddy \& Steidel
(2004), a slope of
$\alpha=-0.8$ is adopted to derive an average 1.4 GHz rest-frame
luminosity of $16\times10^{22}$W~Hz$^{-1}$, corresponding to an
average SFR$\sim160$~$M_\odot$yr$^{-1}$ per object, using the relation
given by Yun et al. (2004) and corrected for a binning error as in Reddy
\& Steidel (2004)\footnote{Reddy, private
communication}.  
Including again into the sample the two
starburst galaxies detected at 1.4 GHz we obtain an average
SFR$\sim270$~$M_\odot$yr$^{-1}$ per object, again with reasonable
consistency with both the optical and the X-ray estimates.  Also in
this case, no statistically significant difference is found for the
average radio flux density of objects with or without known spectroscopic
redshift.

\begin{figure}[ht]
\centering 
\includegraphics[width=8.8cm]{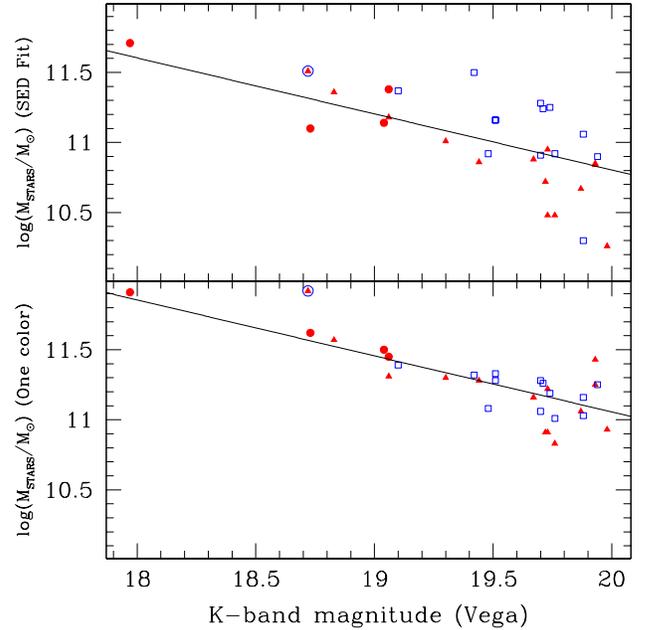}
\caption{The stellar masses for the K20/GOODS objects at $z>1.4$ are shown
as a function of $K$-band magnitudes. The two plots presented refer to each 
of the two methods discussed by Fontana et al. (2004) to estimate the galaxy
stellar masses (see text). Symbols are as in Fig.~\ref{fig:BzK}.  }
\label{fig:MassCal}
\end{figure}

To summarize, all the available SFR indicators agree with each other
and confirm the presence of $K$-band luminous $<\!z\!>\simeq 2$ 
star-forming galaxies with typical SFR$\approx200$~$M_\odot$yr$^{-1}$ 
and a median reddening of $E(B-V)\sim0.4$ (cf. D04).

\section{The Stellar Mass of {K}-Selected Galaxies}
\label{sec:masses}

As part of the K20 project the stellar mass $M_*$ of each galaxy was
estimated from the known redshifts and full multicolor photometry
(Fontana et al. 2004; F04 hereafter).  Using the F04 results, in this
section we explore the possibility of estimating the stellar-mass
content of $K$-selected galaxies at $z>1.4$ from the $BzK$ photometry
alone. 

Fig.~\ref{fig:MassCal} shows the results of two different stellar mass
estimates from F04: one based on synthetic stellar population models
fitting to the whole $UBVRIzJHK$ SED, and one based on fitting just
the $R-K$ color. The latter approach is designed to provide an
estimate of the maximal mass of each galaxy (see F04). 
The masses
estimated with the SED-fit technique are in reasonable agreement
with those for the objects at $1.7<z<2.3$ analyzed in D04.
 
Fig.~\ref{fig:MassCal} shows a plot of $M_*$ from both methods as a
function of the observed $K$ magnitude for the sample of 31 out of the
32 objects with $z>1.4$ in the K20/GOODS sample. One object was
excluded because exhibiting a clearly AGN dominated spectrum.
Best-fitting linear relations between the stellar-mass and the observed
$K$-band flux were obtained, in the form:
\beq
{\rm log}(M_*/10^{11} M_\odot) = -0.4(K^{\rm tot}-K^{11})
\label{eq:M1}
\eeq
where $K^{11}$ is the $K$-band magnitude corresponding on average to
a mass of $10^{11} M_\odot$. For the SED fit and single-color method
we find $K^{11}=19.51$ and $K^{11}=20.14$ (Vega scale), respectively.

It can be noted that at $z>1.4$ the
single-color method yields masses a factor of 1.7 higher, on average, than
the SED-fit technique (see also F04). 
The rms dispersions observed for these relations
are $\sigma(\Delta {\rm log} M_*)=0.25$ and 0.15 for the best fit and single color
method, respectively. 
We searched for further correlations between the residuals in the masses
$\Delta {\rm log} M_*$
as derived from Eq.~\ref{eq:M1} versus the F04 
values, and the
colors available from the $BzK$ photometry. 
No significant trend was noticeable  for the 
masses derived with the single-color method. Instead, 
the residuals in the SED fitting derived masses do positively correlate with the $z-K$
color, with: 

\beq
\Delta{\rm log} M_* = 0.218[(z-K)_{AB}-2.29],
\label{eq:delta}
\eeq
a term that would reduce the rms dispersion to 0.20 if added
to the right-hand side of Eq.~\ref{eq:M1}.

These relations allow to estimate masses with average uncertainties
on single objects of
about 40\% and 60\% relative to the single-color and the SED-fit
method, respectively.  This is an encouragingly good accuracy, given
the large intrinsic differences in the luminosity distance and actual
rest-frame wavelength sampled by the observed $K$-band, for objects
within the $1.4<z<2.5$ range.
Intrinsic differences in the $M/L$ ratio for given magnitudes and/or colors
also contribute to increase the scatter.  However, when averaging over
large samples of galaxies these statistical fluctuations may be
largely  mitigated. 

Note also from Fig.~\ref{fig:MassCal} that the stellar masses of dusty
star-forming and old/passive galaxies are estimated to be
on average quite similar at given  observed $K$-band magnitude. 
This seems to happen by chance: the star-forming galaxies have lower
mass to light ratio but their $K$-band light is attenuated by an
amount which produces similar observed magnitudes to old objects
with comparable stellar-masses. The mass of substantially obscured stellar
populations within the galaxies would however obviously fail to be
accounted for in such estimates.

It should be reminded that relations~\ref{eq:M1} and \ref{eq:delta} 
were derived for $K<20$ galaxies, and it remains to be assessed whether 
they are also valid at fainter $K$ magnitudes.

\section{The B\lowercase{$z$}K vs. Other High-$z$ Galaxy Selection
Criteria}
\label{sec:other}

In this section, the properties of $BzK$-selected galaxies
at $1.4<z<2.5$ having $K<20$ are compared to those of samples selected
according to other color or multi-color selection criteria.
We will consider the $U_{\rm n}GR_{\rm s}$ selection of $z=2$
galaxies, the Extremely Red Objects (ERO) selection based on
the $R-K>5$ threshold,  and the infrared-selected galaxies
found with the criterion $J-K>2.3$ proposed by Franx et al. (2003) 
to isolate $z>2$ evolved galaxies.

\begin{figure}[ht]
\centering 
\includegraphics[width=8cm]{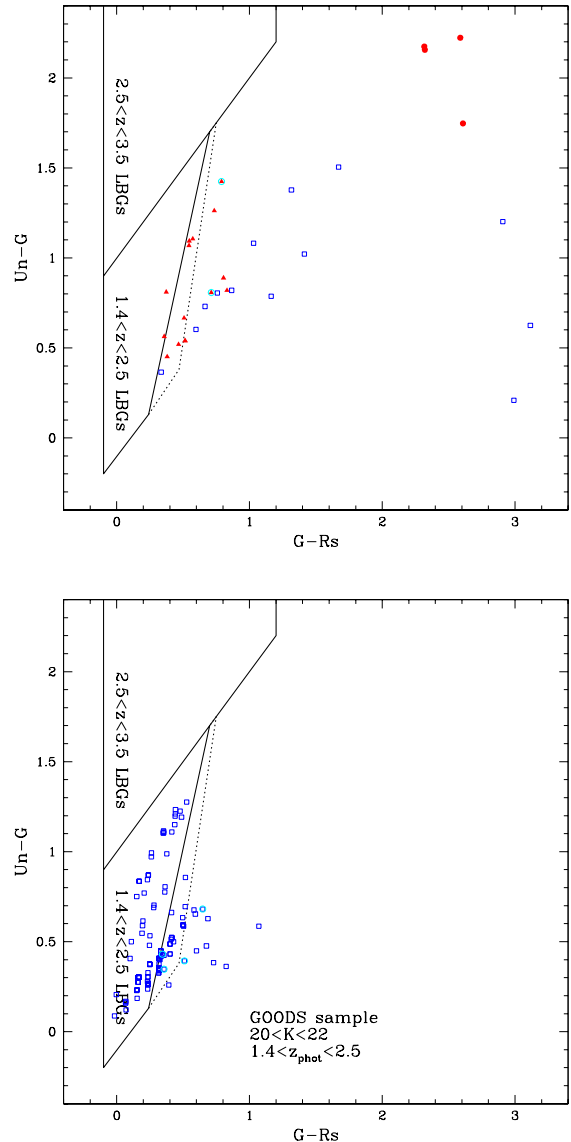}
\caption{The $U_{\rm n}GR_{\rm s}$ two-color diagram for $1.4<z<2.5$ galaxies in
the K20/GOODS region (top panel) that are selected by the criteria defined
in Section ~\ref{sect:BzK}, and (bottom) fainter galaxies with
$1.4<z_{\rm phot}<2.5$ and $20<K<22$ from the same K20/GOODS area (the sample
also shown in Fig.~\ref{fig:IS2}).  The $U_{\rm n}GR_{\rm s}$ colors
were derived from the SED fitting.  Symbols are as in
Fig.~\ref{fig:BzK} (in the bottom panels all redshifts are
photometric).  In both panels, the color regions defined for the
UV identification of $z\sim2$ are also shown (Steidel et al. 2003 for
$z\sim3$; Erb et al. 2003, solid line, and Adelberger et al. 2004,
dotted line, for $z\sim2$).}
\label{fig:LBG_z2}
\end{figure}

\subsection{$BzK$- vs. UV-selected Galaxies at $z\sim2$} 
\label{sec:UV}

Very recently, the UV technique for selecting LBGs has been extended
to $z<3$ using a $U_{\rm n}GR_{\rm s}$ two-color diagram which isolates
star-forming galaxies at $1.4<z<2.5$ (Erb et al. 2003; Steidel et al.
2004; Adelberger et al. 2004), a redshift range fully matching that
of the $BzK$ selection.

\begin{figure}[ht]
\centering 
\includegraphics[width=8.8cm]{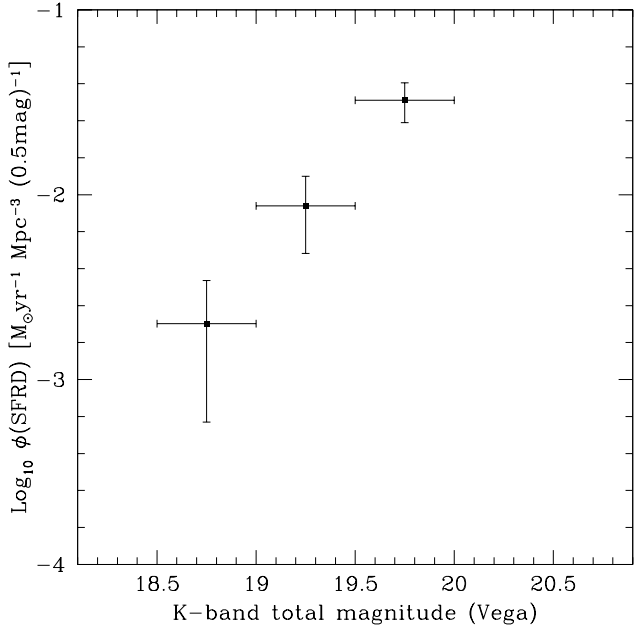}
\caption{
The differential contribution to the  SFR density at $z\simeq 2$ from
$BzK$ galaxies as a function of their $K$-band magnitude (for K20 galaxies
in the K20/GOODS region). 
Note that this contribution is still increasing at the $K=20$ limit of
the survey, suggesting a non-negligible contribution from unaccounted $K>20$
galaxies. Error bars are purely Poissonian.
}
\label{fig:SFRplot}
\end{figure}

The UV-selection requires the UV continuum to be relatively
flat, thus limiting the overall dust extinction 
to $E(B-V)\simlt0.3$ (Adelberger \& Steidel 2000), while many of the K20
galaxies at $z>1.4$ are more reddened (see Fig.~\ref{fig:redde}). 
We estimated how many of the galaxies at $1.4<z<2.5$ could
also be selected by the UV criterion, in our sample. 
As no $U_{\rm n}GR_{\rm s}$ 
photometry is available to us for the K20 galaxies, 
synthetic $U_{\rm n}GR_{\rm s}$ magnitudes have been derived
from the BC03 models providing the best
fit to the observed $UBVRIzJHK$ SEDs, for the objects in the K20/GOODS
region. Fig.~\ref{fig:LBG_z2} shows the resulting synthetic
$(G-Rs)$ vs. $(U_n-G)$ colors. 
Of the 32 $K<20$ objects at $1.4<z<2.5$ in Fig. 3, 
only 2 (6\%) would be selected by
the Erb et al. (2003) criteria, and 9 (28\%) by the Adelberger et al. (2004)
criteria, corresponding to a surface density of about
0.3 arcmin$^{-2}$. We can compare these numbers  with those in
the UV-selected surveys. The Adelberger et al. (2004) BM and BX criteria 
result cumulatively in a sky density of 9 arcmin$^{-2}$ candidate
$z\sim2$ galaxies (Steidel et al. 2004). Roughly 90\% of these is
found within $1.4<z<2.5$ and
about 8\% of the $z\approx2$ objects has $K<20$. Based on the above
numbers, we should have found $\sim20$ galaxies UV-selectable with 
the Adelberger et al. (2004) criterion in
$1.4<z<2.5$ in the K20/GOODS region, while only 9 are recovered.
It is not clear what is the reason of this possible discrepancy, that
may be in part due to small number statistics
and/or to cosmic variance due to clustering (D04).

Some large fraction (perhaps as high as $\sim 70\%$) of the $K<20$ 
galaxies at $z>1.4$ fail to be
selected by the UV criteria to identify galaxies at $z\sim2$. The
lost fraction includes not only the old passive systems but also a
high proportion of actively star-forming, highly reddened galaxies. 
As a consequence,
the UV-selection fails to recover most of the stellar mass in $K<20$
galaxies at $z=2$, as expected given that it was not devised with the
aim of probing the galaxy mass density, but rather the star-formation rate
(Adelberger et al. 2004). However, a significant amount of the 
SFR density is also missed. Using the SFRs
estimated from reddening-corrected UV luminosities
(Section~\ref{sect:ebv}), it is found that the 9 $K<20$ starbursts
in our sample
that satisfy the $U_{\rm n}GR_{\rm s}$ criteria
produce only $\approx 15$\% of the SFR density at $z=2$
from $K<20$ galaxies, with
the residual $\approx 85\%$ being missed because of dust reddening in excess of
$E(B-V)\approx0.3$.
The missed objects include all
the K20 galaxies with the most extreme starbursts with SFR$>200
M_\odot$ yr$^{-1}$.

Comparing the typical star-formation rate level per galaxy, the bright
$K<20$ starbursts in the K20 survey appear to be forming stars more
vigorously than the UV-selected not strongly reddened
galaxies with $K<20$ (Shapley et al. 2004), with average SFRs larger by
a factor of $\approx3$. Extending the comparison including also 
fainter UV selected galaxies,
the radio and X-ray measurements (both in
agreement with the extinction-corrected estimates based on the UV 
continuum), imply that the average radio and X-ray luminosities, hence the 
SFRs,
of our $K<20$ $z=2$ star-forming objects are higher than those of 
the average of all UV selected galaxies
(Reddy \& Steidel 2004) by a factor of $\sim 4$. 
The $U_{\rm n}GR_{\rm s}$ selected galaxies
studied by Reddy \& Steidel (2004) have a
significantly higher space density than the $K<20$ galaxies and sample 
regimes with much lower SFRs.

\subsection{Contributions to the $z\sim2$ Star-formation Rate Density}

In this section we derive the contribution of the $BzK$-selected
galaxies to the {\rm integrated} star formation rate density (SFRD)
at $z\sim2$, and compare it to an estimate of the SFRD derived from
the UV-selected galaxies.

For the volume in the redshift range
$1.4<z<2.5$, a SFRD of $0.044\pm0.008$ $M_\sun$
yr$^{-1}$ Mpc$^{-3}$ is derived from the 24 K20 star-forming galaxies 
fulfilling the $BzK>-0.2$ criterion (and of course $K<20$), where the error
is derived from bootstrap resampling. This may well be an underestimate of 
the error, given the 
small number of galaxies used in the estimate, and to them belonging
to a population likely to be strongly
clustered (D04). For example, assuming that these galaxies  are as
clustered as $z\sim1$--3 red galaxies ($r_0\simlt10$ \h1 Mpc; e.g., Daddi et
al.\ 2001; 2003), the error would become 35\% larger on the
lower side and a factor of 3 larger on the upper-side.

This estimate of the SFRD contributed by the $BzK>-0.2$ objects with
$K<20$ is comparable to the global SFRD at $z\approx2$ as estimated 
from other surveys (in the same units: $\sim0.08$, Connolly et
al. 1997, corrected for extinction by Steidel et al. 1999; 
$\sim0.055$, Heavens et al. 2004), or as predicted by $\Lambda$CDM 
semi-analytical models (0.05--0.10 $M_\sun$ yr$^{-1}$ Mpc$^{-3}$, 
Somerville et al. 2001) and by  $\Lambda$CDM hydrodynamical 
simulations ($\sim0.055$ $M_\sun$ yr$^{-1}$ Mpc$^{-3}$, Hernquist \& 
Springel 2003). 
 However, our present estimate must be incomplete because it
 does not include the contributions of all the $K>20$ galaxies, and
in particular of those still fulfilling the $BzK>-0.2$
condition. Fig.\ref{fig:SFRplot} shows that the SFRD is
 not yet converging by $K\sim20$, and a significant additional
 contribution from $K>20$ galaxies is therefore expected.
 
The {\rm total} SFRD produced by UV-selected galaxies at $z\sim2$ in
the Steidel et al. (2004) sample has not been published yet, but a
crude estimate can be derived in comparison to the $BzK$ star-forming
galaxies at
$K<20$ by considering that the $z\sim 2$ $U_{\rm n}GR_{\rm s}$-selected
candidates down to $R_{\rm s}=25.5$ have
$\approx10$ times higher sky density (Steidel et al. 2004), that $\sim
90$\% of them are in the redshift range covered by the $BzK$
selection ($1.4<z<2.5$), and that they
have $\sim 4$ times smaller average SFRs. This would yield an
integrated contribution by UV-selected galaxies (with $R_{\rm s}<25.5$)
a factor of 2--2.5 times larger than that of bright $K<20$ $BzK$ galaxies.
Hence, taking into account that some galaxies are picked
by both criteria, the UV selection may miss of order of $\sim 20$--30\%
of the total SFRD provided by galaxies selected by at least one the
two criteria, while the $BzK$ criterion limited to $K<20$ allows to
select only a similar fraction.

On a broader perspective, discussing the potentials of the two color
criteria ($BzK$ and $U_{\rm n}GR_{\rm s}$) is perhaps more interesting
than comparing the existing samples drawn with them.  As mentioned
above, much of the limitation of the present application of the $BzK$
criterion actually comes from the fairly bright limiting $K$
magnitude, rather than from the color criterion itself. One may expect
that applying it to fainter $K$ magnitudes a higher fraction of the
total SFRD could be recovered. It is not presently known, however, if
some fraction of $z=2$ galaxies would be missed by the $BzK$ criterion, 
especially at $K>20$, where it may start
loosing some very young starburst, as suggested by the top-left panel
in Fig. 8. If they exist at faint $K$ magnitudes, such young galaxies 
could be more easily selectable in the UV.
These points should be tested by future surveys\footnote{We are grateful to the referee (C. Steidel) for having
informed us prior of publication that in his sample of $z\sim2$ galaxies
(c.f. Sect.~\ref{sec:UV}) the $BzK>-0.2$ selection appears to miss a
fraction of star-forming galaxies with $K>21$.}.

The UV-selection 
appears to miss the most actively star-forming galaxies not because 
of the limiting $R$ magnitude, but because they are
much too reddened [$E(B-V)\simgt0.3$] for satisfying the $U_{\rm
n}GR_{\rm s}$ color selection. The existence of highly reddened
star-forming galaxies also at $K>20$ would imply for the
UV-selection additional losses of star-forming galaxies (hence of part
of the SFRD).
A preliminary analysis based on the
GOODS/ISAAC sample at $K<22$ (Fig.~\ref{fig:IS2}) suggests that at
$K>20$ star-forming galaxies with $z_{phot}\sim2$ and progressively bluer
colors start to appear, and occupy the bluest $U_{\rm n}GR_{\rm s}$
region where most UV-selected galaxies also lie (Fig.~\ref{fig:LBG_z2}, 
see also Fig.~11 of Adelberger et al. 2004 for comparison). This may be 
consistent with a general blueing trend
at fainter magnitudes, indicative of a trend to lower reddening. The
fractional SFRD lost by the UV selection may then therefore 
decrease with increasing $K$ magnitude limit.  It should be
mentioned, however, that the paucity of red star-forming (as well as of
passive) galaxies in the sample of Fig.~\ref{fig:IS2} is at least in
part due to the requirement of accurate photometry for the
photometric redshifts determination and that, e.g., candidate
star-forming galaxies with $BzK>-0.2$, $z_{\rm phot}\approx2$, and
$2<(z-K)_{AB}<4$ are found also down to $K=22$.  In addition, a
population of faint red galaxies at $z\simgt2$ appear to exist even down to
$K=24$. They tend to be more clustered than LBGs, as expected
for the precursors of early-type galaxies (Daddi et al. 2003).  In
summary, an application of the $BzK$ technique to much fainter
$K$-selected surveys could shed light on the amount of reddened
star-formation at rates lower than probed by the K20 survey, and
better establish the fractions of the SFRD recovered by each of the
two criteria.

\subsection{$BzK$-selected Galaxies and Extremely Red Objects}

A simple method to select relatively high redshift galaxies relies on
requiring very red optical to near-IR colors, typically $R-K>5$
(e.g. Elston, Rieke \& Rieke 1988; Hu \& Ridgway 1994; Thompson et
al. 1999; Daddi et al. 2000a; Roche et al. 2002, 2003; for a comprehensive
review see McCarthy 2004). K20 survey
spectroscopy unveiled for the first time the nature of EROs in 
a sizable sample, and showed that EROs include  similar fractions
of old and dusty star-forming systems (Cimatti et al. 2002a, 2003; 
see also Yan et al. 2004).

\begin{figure}[ht]
\centering 
\includegraphics[width=8cm]{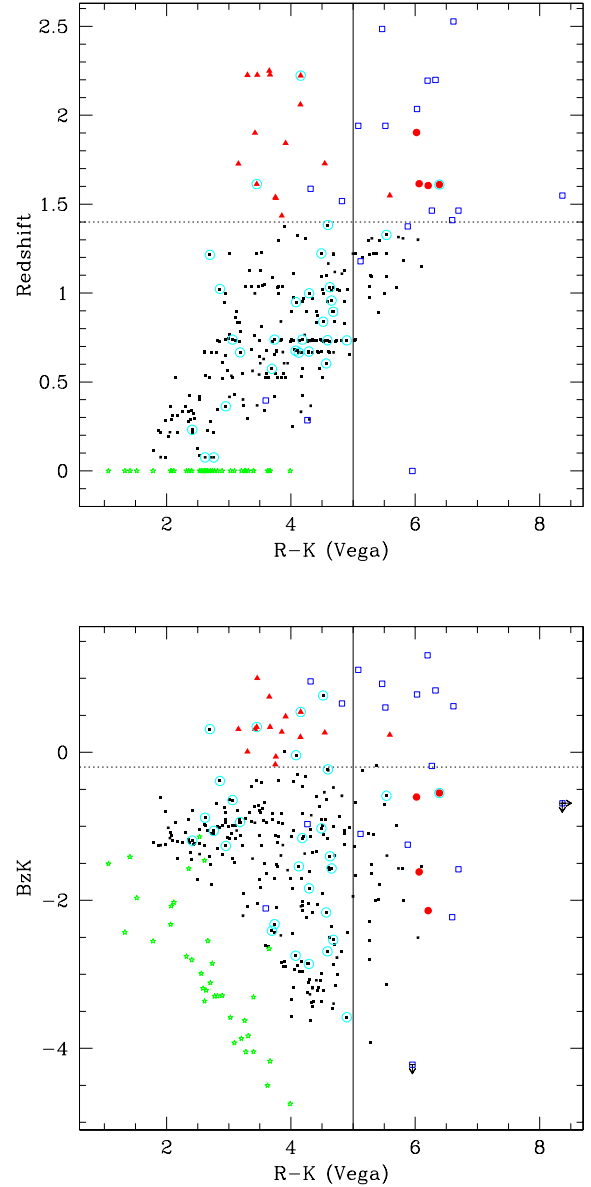}
\caption{The $R-K$ (Vega) color versus redshift (top)
and versus $BzK$ (bottom), for galaxies in the K20/GOODS region.
Symbols are as in Fig.~\ref{fig:BzKvsz}.
}
\label{fig:EROS2}
\end{figure}

The redshift range is one of the main differences among the samples
produced with the two $BzK$ and ERO methods. Fig.~\ref{fig:EROS2} 
shows that EROs with $K<20$ are found at $0.8\simlt z\simlt2.5$, coherently with the
rationale for their selection (e.g. Daddi et al. 2000a). Many K20 galaxies
exist in the same redshift range that are not EROs. In contrast, the
strength of the $BzK$ selection (Eq.~\ref{eq:cond1} and \ref{eq:ES0}) 
is that it provides a fairly
complete sample of galaxies in the redshift range
$1.4<z<2.5$. The $z-K>2.5$ condition of Eq.~\ref{eq:ES0} that allows to
recover old passive galaxies is basically equivalent to the ERO
criterion $R-K>5$, apart from the higher low-redshift cutoff
($z\simgt1.4$ instead of $\simgt 0.8)$.  A very similar $z>1.4$ cutoff
for passive galaxies
would be obtained requiring $R-K\simgt6$, albeit with a larger
contamination by both lower redshift and reddened star-forming galaxies.  
Fig.~\ref{fig:EROS2} shows
that only about 50\% of $BzK$ selected galaxies at $z>1.4$ have EROs
colors, while only 35\% of all the EROs are selected with the $BzK$
criteria, i.e. lie at $z>1.4$.

\begin{figure}[ht]
\centering 
\includegraphics[width=8cm]{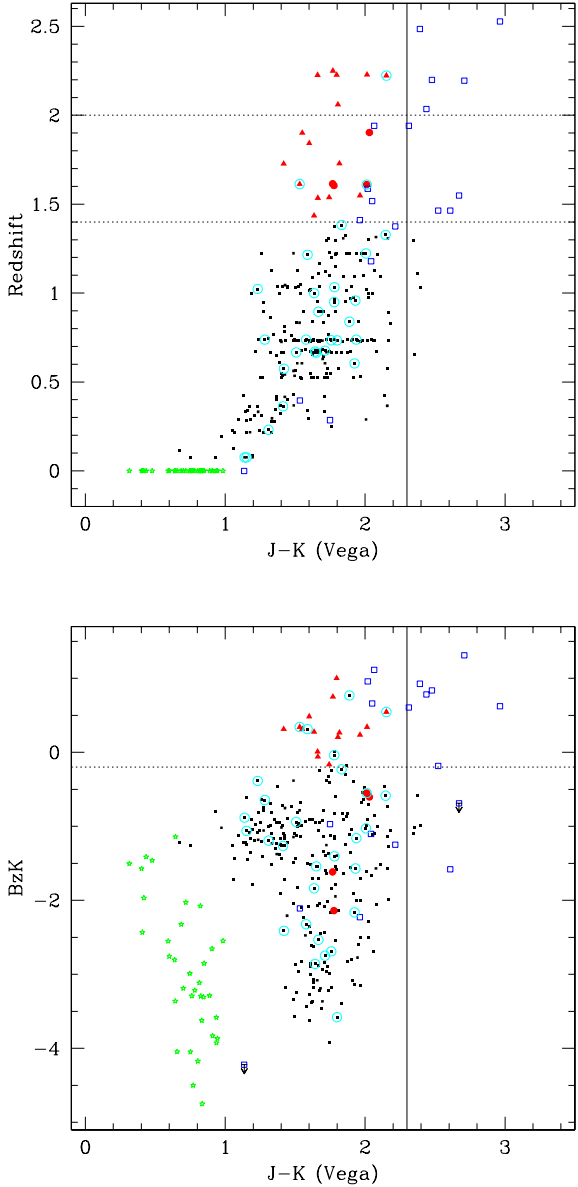}
\caption{The $J-K$ (Vega) color versus redshift (top)
and versus $BzK$ (bottom), for galaxies in the K20/GOODS region.
Symbols are as in Fig.~\ref{fig:BzKvsz}.
}
\label{fig:JK2}
\end{figure}

Particularly interesting is the comparison of the physical properties
of dusty star-forming EROs to those of $z\sim2$ starbursts. The two
samples include a few common objects, but it appears that on average
the SFR of the $z=2$ star-forming galaxies is one order of magnitude
higher than that of star-forming EROs.  Indeed, the average X-ray 2--10 keV
luminosity of dusty EROs in the K20 survey with $<\!z_{\rm spec}\!>
=1.053$ (Brusa et al. 2002) is a factor of $\sim10$ smaller than that
measured for the $z=2$ $BzK$-selected starbursts. Similarly, the
average 1.4 GHz luminosity of the same EROs is a factor of $\sim6$
smaller than found at $z=2$ (Cimatti et al. 2003).  As these estimates
are limited to EROs with known spectroscopic redshift, they exclude the
highest redshift $z_{\rm phot}\sim2$ EROs in common with the $BzK$ selected
sample. 
Forming stars much more vigorously, the reddened
starburst galaxies seen at $z\sim2$ appear therefore
to be of a different nature 
with respect to dust-reddened galaxies at $z\sim1$.

Similarly to the method proposed by Pozzetti \& Mannucci (2000; PM2000 hereafter),
requiring $BzK\geq-0.2$ would allow to distinguish dusty EROs from the
old ones, extending the diagnostic in $1.4<z<2.5$ (the PM2000
criterion is formally valid only up to $z\sim2$).  
The two criteria are however substantially different
and complementary to each other: while the PM2000 criterion relies on detecting the
signature of the 4000 \AA\ break of old galaxies, the $BzK$ criterion
aims at detecting the UV tail in the SEDs due to the
youngest stars, even in the presence of substantial reddening.  
We verified that 8/9 of the EROs that are also
star-forming galaxies at $z>1.4$ with $BzK>-0.2$, mostly objects with
photometric redshifts only, are correctly classified as star-forming
galaxies by the PM2000 criterion.  

It would be extremely interesting to apply the $BzK$ diagnostic to 
EROs samples in order to statistically distinguish $z<1.4$ EROs from those
at $z\simgt1.4$, either old or star-forming ones. This would allow to solve
the long standing issue of whether the EROs overdensities observed in the 
field of AGN/QSO at $z\simgt1.5$ are true spatial associations (i.e. 
clusters or proto-clusters) or are just due to lensing effects (e.g., Cimatti et al. 2000; 
Best et al.  2003; Wold et al. 2003). 
In case of true spatial associations one would also
know if such enhancements are due to dusty star-forming
or passive galaxies with 
important implications for galaxy formation in clusters.

Finally, we notice that the color properties of star-forming galaxies at
$z\sim2$, having blue $B-z$ colors and the reddest $z-K$ color, are fully
matching those of the mysterious population of {\em red outlier}
galaxies (Moustakas et al. 1997). 
These galaxies were found to show blue
$V-I$ colors with respect to their large $I-K>4$ colors and their nature had
remained so far unclear.

\subsection{$BzK$- vs. $(J-K)$-selected Galaxies}

Recently, Franx et al. (2003) have proposed the criterion $J-K>2.3$ to select
evolved galaxies at $z>2$, and spectroscopic evidence that $z>2$
galaxies are indeed selected by this criterion has been provided for a
sample of star-forming galaxies and AGN
(van Dokkum et al. 2003; 2004).  Based on a
single color, the $J-K>2.3$  criterion  is similar to that for EROs
($R-K>5$), and like this one it may eventually result to select both reddened
star-forming galaxies and evolved/passive ones.
The properties of $J-K>2.3$ galaxies in the K20 survey were then
examined and compared with those recovered by the $BzK$ criteria.
Fig.~\ref{fig:JK2} shows the $BzK$ vs $J-K$ diagram 
and the redshift vs $J-K$ diagram for galaxies in the K20/GOODS region.  
Only 4 galaxies
with $J-K>2.3$ have a spectroscopic redshift, and lie in the range
$0.6<z<1.3$. Additional 3 galaxies have a photometric redshift
$z\sim1.5$, and only 6 out of 13, of the $J-K>2.3$ objects, have
$z_{\rm phot}\simgt2$. Therefore, it seems that the contamination of $z<2$
galaxies among $J-K$ red galaxies could be higher than in the van 
Dokkum et al. (2003)
sample. In part this could be due to photometric errors, as the
lowest $z$ contaminants have $2.3<J-K<2.5$, hence close to the edge
of the $J-K>2.3$ region. Note however that both the van Dokkum et al.
spectroscopic sample and the K20/GOODS $J-K>2.3$ sample are quite small.

Only 9 out of 32 objects in the K20/GOODS sample at $1.4<z<2.5$ fulfill the
$J-K>2.3$ condition, but the fraction rises to 5/11 for the $z>2$ galaxies.
All the galaxies having $J-K>2.3$ and $z_{phot}>1.9$ have also $BzK>-0.2$,
and would be classified as reddened star-forming galaxies rather than
purely passive systems. This agrees with the recent results by van Dokkum et
al. (2004) and F\"orster Schreiber et al. (2004).
The highest redshift passive systems at $1.6<z<2$
in the K20 survey (Cimatti et al. 2004) have $J-K\sim1.7-2$.

In summary, it appears that the $BzK$ selection has the advantage of
allowing in principle to recover the bulk of the galaxy population for the
redshift range $1.4<z<2.5$ for which it is tuned, 
including the reddest and bluest ones, and to distinguish the passive
from the star-forming ones.
However, while the $BzK$ selection is efficient only up to $z\sim2.5$,
the $J-K>2.3$ criterion can allow to pick up the reddest galaxies up to much 
higher redshifts $z\simlt4$ (Franx et al. 2003).

\begin{figure}[ht]
\centering
\includegraphics[width=8.8cm]{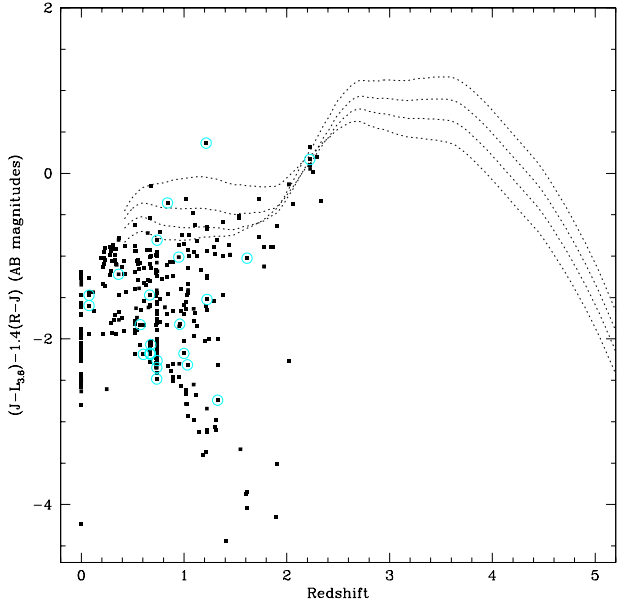}
\caption{A possible reddening independent
selection criterion for $2.5\simlt z\simlt 4.0$  star-forming galaxies
is obtained with $RJL\equiv J-L_{3.6}-1.4(R-J)>0$ (AB magnitudes). The
above quantity is plotted for the galaxies in the K20 survey
(the $L$-band magnitudes were derived from the best-fitting
SED) and for constant
star-formation rate models and 0.2, 0.5, 1 and 2 Gyr ages.
Circled points show X-ray detected sources.
}
\label{fig:RJL}
\end{figure}

We finally notice that most of $BzK$ galaxies have $J-K>1.7$
(Fig.~\ref{fig:JK2}). At this
threshold the clustering of faint $K$-selected galaxies at $z\simgt2$
was observed to become quite strong, compared to bluer galaxies (Daddi
et al. 2003). This is consistent with $BzK$-like galaxies contributing
to such a clustering enhancement (as suggested by D04), 
together with the reddest $J-K>2.3$ galaxies.

\section{Extending the Technique to Select $\lowercase{z}>2.5$ Galaxies
with Spitzer Photometry}
\label{sec:SIRTF}

The $BzK$ criterion is based on the rest-frame colors of (reddened)
star-forming  and (unreddened) passive galaxies, and then tuned to
select those at $z\sim2$. Therefore, by choosing a different set of
bands that sample the same rest-frame wavelengths one can forge a new
criterion tuned to select the same kind of galaxies at a higher redshift.

By multiplying by 1.5 the central wavelengths of the $BzK$ bands
one obtains values that roughly correspond to the central wavelengths
of the $RJL$ bands, and correspondingly a criterion based on the
quantity:

\beq
RJL \equiv (J-L_{3.6})_{AB} - 1.4(R-J)_{AB}
\label{eq:RJL}
\eeq
\smallskip
\noindent
can be used to select galaxies in the redshift range $2.5\simlt
z\simlt 4$, i.e., complementary to the $BzK$ criterion that can select
galaxies up to $z\sim 2.5$.
The $RJL$ quantity has its
peak for star-forming galaxies in the above redshift interval, again
due to the Balmer break being located between the $J$ and 
$L$-band at 3.6 $\mu$m. 
Given that the ratio of the central wavelengths of the $RJL$ bands 
to the $BzK$ ones is not exactly constant, 
a factor $1.4$ in Eq.~\ref{eq:RJL} is necessary to
make $RJL$ reddening independent in $2.5<z<4$ (using the reddening law of 
Calzetti et al. 2000).
Model tracks for the $1.4(R-J)_{AB}$ versus 
$(J-L_{3.6})_{AB}$ colors in the range $2.5<z<4.0$ are nearly
identical to those for the $(B-z)_{AB}$ versus $(z-K)_{AB}$ colors 
in $1.4<z<2.5$, already discussed in Section ~\ref{sec:modeling} 
and shown in the panels of Fig.~\ref{fig:4p}.
By requiring $RJL\simgt 0$ one should thus in
principle cull $z\sim2.5$--4 star-forming galaxies in $L$-band limited
samples, independently on their reddening, while objects having
$RJL<0$  and $J-L_{3.6}\simgt2$--2.5 should turn out to be passive 
objects at $z>2.5$ (if such a population of galaxies exists).
As a consistency check, synthetic $L$-band
magnitudes were extrapolated for K20 galaxies from their best fitting 
SEDs to test for contamination by $z<2$ galaxies in $RJL\simgt 0$
selected samples, that results to
be small (Fig.~\ref{fig:RJL}), though the K20 sample does not cover
the redshift range $z>2.5$ that should be sampled by the  $RJL>0$ criterion.
A few  $z>2$ galaxies in the K20 sample
start to show $RJL>0$ following the models trend.
A possible problem of this $RJL$ selection technique might be contamination by
low redshift (e.g. $z\simlt0.5$) galaxies that could arise if significant
contribution by dust (e.g. from AGN) is starting to appear in the
$L$-band. No contribution of this kind was considered in the derivation
of $K20$ synthetic $L$-band magnitudes. 
It is unclear how often this can happen for faint
low-redshift galaxies. Filtering of low-redshift interlopers may be
desirable for the application of this technique if the above
contamination should result to be relevant.

If placed at $z=3.5$ a typical $z=2$ star-forming galaxy in the
K20 survey would have $L_{3.6}\sim 22$--23 (AB). This is much brighter than the
limits that should have been  reached by the GOODS-SST
observations at 3.6 $\mu$m
(Dickinson et al. 2002). As an example, for $E(B-V)\sim0.6$ at $z\sim3.5$ 
one expects colors $(R-L_{3.6})_{AB}\approx5$ and
$(J-L_{3.6})_{AB}\approx3$, implying the need to reach quite faint magnitudes
in the optical/near-IR in order to detect
such galaxies, i.e., $R\sim27$--28 and $J\sim25$--26 (AB scale
magnitudes).
A first check of this criterion should be possible with the
GOODS ACS+ISAAC+SST dataset, that is expected to be deep enough in all the 
$RJL$ bands. In particular, it will be possible to test whether galaxies
exist in the range $z\approx 2.5-4.0$ that are picked up by the $RJL$ selection
but  missed by the $U_{\rm n}GR_{\rm s}$ UV-selection, in analogy to
what found for the $BzK$ selection.

\section{Discussion}
\label{sec:discussion}

\subsection{Early-Type Galaxies in Formation}

The masses of the $z\sim2$ $BzK$ galaxies in the K20 survey are
overall quite high, with a median of $\sim 10^{11} M_\odot$, and in
the local universe objects with such high masses are almost uniquely
found among early-type galaxies.
In D04 we had in fact suggested that the  properties of
$K$-selected galaxies at $z=2$ are consistent with those expected for
the star-forming precursors of massive spheroids.
This was also supported by the high SFRs, 
large sizes,
merging-like morphologies and strong redshift space spikes
that hint for strong clustering. Additional evidence in this direction
comes from their strong photospheric and interstellar lines,  indicative 
of solar metallicity or above, typical of massive spheroids
(de Mello et al. 2004). 

The properties of the $BzK$-selected galaxies with $K<20$ suggest that there is
a whole population of vigorous starbursts  within the $1.4<z<2.5$ range that
can qualify as spheroids in the making. Such properties 
are quite different from those of  star-forming galaxies at lower redshifts, 
e.g., the SFRs are $\sim 10$ times higher than those of  $z\sim1$ dusty EROs, 
which in turn may be much less strongly
clustered than $BzK$-selected galaxies (Daddi et al. 2002, 2004).

High redshift dusty star-forming galaxies selected at submm/mm
wavelengths (see Blain et al. 2002 for a review) are often considered as
the precursors of the present-day massive spheroids. Although a
detailed comparison between $BzK$-selected and submm/mm-selected
galaxies is beyond the scope of this paper, we notice that the latter 
systems have a redshift distribution largely overlapping with that
of the former ones (median redshift of $z\sim2.4$; Chapman et
al. 2003), even higher SFRs (Blain et al. 2002), similarly high masses (Genzel et al.
2003) and possibly  similarly high clustering (Blain et al. 2004). 
However, the space density of submm/mm-selected galaxies is
a factor of 10--100 lower than that of the $BzK$-selected ones.
Submm/mm sources may be extreme subsets of $BzK$ galaxies.

\subsection{Entering the Spheroid-Formation Epoch? A V/V$_{\rm max}$ Test}
\label{sec:vmax}

Many lines of evidence suggest that ellipticals and bulges formed the
bulk of their stars at $z\simgt 2.5-3$, both in clusters (e.g., Bower,
Lucey, \& Ellis 1992) and in the field (Bernardi et al. 1998, 2003),
with much evidence having been accumulated from both low- and
high-redshift ($z\sim 1$) observations of passively evolving spheroids
(see e.g., Renzini 1999 for an extensive review; see also
Thomas et al. 2002). 
This is further
reinforced by the recent discovery of passive early-type galaxies
at $z\sim2$, with UV-luminosity-weighted ages of 1--2 Gyr, implying
formation redshifts beyond $z\simgt 2.5$--4 (Cimatti et al. 2004).
Therefore, a natural question is whether the $z\sim 2$ $BzK$-selected
starbursts belong to the major epoch or to the low redshift tail of
spheroid formation.

In order to tentatively distinguish between these alternatives, a
$V/V_{max}$ test (Schmidt 1968) was performed on the flux limited
population of $K$-selected star-forming galaxies at $z>1.4$.  For each
object, $V/V_{\rm max}$ is computed as the ratio between the volume
within the range $1.4<z<z_{obj}$ and that within $1.4<z<z_{max}$,
where $z_{\rm max}$ is the maximum redshift for which the object would
still be detected with $K<20$. To compute $z_{\rm max}$ we use the
observed $J-K$ color of each galaxy to estimate the $K$-correction.
In the case of a non evolving population the distribution of $V/V_{\rm
max}$ values should have an average of 0.5. For the 24 star-forming
galaxies with $z>1.4$ in the K20/GOODS region
we derive $<\!V/V_{\rm max}\!>\ = 0.594\pm0.048$. 
Considering only the objects in $1.4<z<2.5$, thus
limiting $z_{\rm max}<2.5$, results in $<\!V/V_{\rm max}\!>\ =
0.64\pm0.05$.
A $<\!V/V_{\rm max}\!>$
greater than 0.5 suggests that the comoving number density of these galaxies is
increasing with redshift, but given the small sample the effect is
only at the 2--3$\sigma$ level. Limits of these calculations are also
that they are based in part on the use of photometric redshifts and that
the results could be well affected by cosmic variance due to the clustering
(D04).

The significance and rate of evolution could be however higher than
recovered here, because biases are likely
to work against the detection of objects at higher and higher
redshifts.  For example, the above calculation assumes no intrinsic
evolution in the luminosity of the galaxies. On the other hand, if objects at
lower redshift (e.g., at $z<2$) were forming stars with similar rates
also at higher redshift (e.g. $z>2$), then the former would be
intrinsically more luminous than the latter ones, and the amount of
evolution would be higher than estimated above. Moreover, the strong
bias due to surface brightness dimming with increasing redshift was
not considered. Given the typical large sizes and low surface
brightness of many of these star-forming galaxies (D04;
Fig.~\ref{fig:morphoBzK}) one expects this effect to be relevant and
to bias to low values the $V/V_{max}$ estimates. 

Consideration of
these effects would further enhance the significance and amplitude of
the evolution, suggesting that by $z\sim 1.4$ we may have just started 
entering the epoch of widespread starburst activity, i.e., of major 
formation of galactic spheroids.
An application of the $V/V_{max}$ test with upcoming larger 
redshift surveys should shed more light on this important point.

\subsection{The Space  Density of Vigorous $z\sim2$ Starbursts:
Comparison with  Models}

We have shown that a substantial population of vigorous starburst
galaxies with average $SFR\approx200$ $M_\odot$ yr$^{-1}$ exists
at $1.4<z<2.5$.
The number density of such $BzK$-selected starbursts can be further compared
to  predictions of theoretical galaxy formation models.

For example, the GIF semi-analytical models (Kauffmann et al. 1999;
Kaviani, Haehnelt \& Kauffmann 2003) predict that within $1.5\simlt z\simlt2$ 
the population of galaxies with masses in the range $10.5<{\rm
log}(M/M_\odot)<11.3$ (similar to the one derived for the K20 $z>1.4$
objects) are either passive galaxies with no ongoing SF or very active
starbursts with SFR$\simgt50$ $M_\sun$ yr$^{-1}$.  This is in
very good agreement with our observations, but in these models the
number density of objects with SFR$>100 M_\sun$ yr$^{-1}$ is
$\sim0.8\times10^{-5}$~Mpc$^{-3}$ and
$\sim1.3\times10^{-5}$~Mpc$^{-3}$ respectively at $z=1.46$ and
$z=2.12$, which is a factor of 10--20 below the observed number
densities.  The space density of passive and massive galaxies is also
similarly underpredicted.

Somerville et al. (2004) provide a mock catalog of galaxies with
$K<20$ based on an updated semi-analytical model with enhanced
starburst activity. While predicting the highest number density of
$z>1.4$ galaxies with SFR$>100 M_\sun$ yr$^{-1}$ compared to all other
models of this class, still it falls short by about an order of
magnitude with respect to the present findings. Also the space density
of passively evolving galaxies with $K<20$ at $z>1.4$ appears to be
underpredicted by a similar factor by this model (Cimatti et
al. 2004).

In general, $\Lambda$CDM semi-analytical models fail to account for
the sheer number of $z\sim2$ galaxies with $K<20$ (Cimatti et al.
2002c; D04; Somerville et al. 2004).  An exception is the hierarchical
model by Granato et al (2004) based on the assumption of a coeval
growth of QSOs and spheroids, which succeeds in producing the high
space density of near-IR bright $z=2$ galaxies (see also Silva et
al. 2004). However, in its current realization this model predicts
that the $z>1.4$ tail of $K$-selected galaxies is predominantly
populated by passively-evolving spheroids (see Fig. 8 in Silva et
al. 2004), at variance with the observed prevalence of vigorous
starbursts.

The recent $\Lambda$CDM hydrodynamical simulations  by Nagamine 
et al. (2004) appear
instead quite successful in reproducing the space density of 
$M>10^{11}M_\odot$ massive galaxies at $z=2$, as observed in the K20
survey, at least in two out of three of its different realizations. 
In one of their three simulation sets the authors also recover 2 galaxies with
old stellar populations and red $G-Rs$ colors, consistent with the colors
of passive spheroids at $z>1.4$ found in K20 (Fig.~\ref{fig:LBG_z2})
and well matching to the K20 space density of passive sources when
accounting for the different volumes. It is not clear instead if these
models can reproduce the observed high density of vigorous starbursts
with SFR$>100 M_\sun$ yr$^{-1}$, as observed in our survey. 

As recently pointed out by Cimatti et al. (2004) and Grazebrook et al. (2004),
in the traditional semianalytic models
the formation of massive spheroids appears to be delayed to much too low
redshifts. On the other hand, the Granato et al. (2004) models appear
to move in the right direction, by pushing the formation to higher
redshift with strong bursts of star-formation, then quenched by strong
AGN activity.  However, in doing so
they may exceed somewhat, as compared to our findings they appear to
underpredict the number of starbursting galaxies still present at
$z\sim 2$. The present results, and an application of the
$BzK$ selection to substantially larger samples,  may help in further 
tuning theoretical
models toward a more realistic description of galaxy formation and
evolution.

\bigskip

\section{Summary and Conclusions}
\label{sec:summary}

$\bullet$ We have introduced a new criterion for selecting 
galaxies within the redshift range $1.4\simlt z\simlt 2.5$ which is based on
the $BzK$ photometry and allows to identify both active star-forming
as well as passively-evolving galaxies, and to distinguish between the
two classes.  The criterion has been tested empirically -- using the
spectroscopic redshifts and spectral types from the K20 survey
($K<20$) including 32 $z>1.4$ objects out of 504 with a spectroscopic
redshift  -- and justified by simulations showing that active and
passive synthetic stellar populations actually follow this selection
criterion and are correctly identified. Albeit smaller in size, other
spectroscopic samples such as the GDDS and photometric redshift of faint
galaxies from the GOODS samples (as currently
available) confirm that the criterion is effective in
selecting galaxies in the mentioned redshift range and also for limiting
$K$-band magnitudes somewhat fainter than $K=20$. 
We have shown that this $BzK$ criterion provides a very efficient way of 
selecting galaxies at $z\approx2$,
that is not biased against passive galaxies and star forming galaxies
that are highly reddened.

$\bullet$
The classification of $K<20$, $z>1.4$
galaxies as actively star forming or passive
was then complemented by HST/ACS morphologies from the GOODS database,
showing that indeed the spectral and morphological classifications
are generally consistent: star-forming galaxies show clumpy, asymmetric
morphologies typical of starbursts and mergers, while passive galaxies
show symmetric surface brightness distribution in general typical
of early-type galaxies.

$\bullet$ 
It is shown that the $BzK$ photometry can be used to
estimate the internal reddening for the K20 galaxies
classified as
star-forming, and their intrinsic luminosity at 1500 \AA. This
allows an estimate of their dust-extinction corrected SFRs. The X-Ray and
radio luminosities of these galaxies provide SFR estimates in very
good agreement with the ones from the de-reddened 1500 \AA\ luminosity.

$\bullet$ 
A significant population of $z=2$ galaxies with $K<20$, average
SFR$\sim200$ $M_\odot$ yr$^{-1}$, and median reddening $E(B-V)\sim0.4$
is uncovered as a result, with a
high volume density of $\sim10^{-4}$ Mpc$^{-3}$ and sky density 
of $\sim 1$ arcmin$^{-2}$. These vigorous
starbursts produce a SFRD of  $\sim 0.044$ $M_\odot$ yr$^{-1}$
Mpc$^{-3}$, representing a sizeable fraction of the total SFRD at $z=2$
as currently estimated.

$\bullet$
For $BzK$-selected galaxies at $1.4\simlt z\simlt 2.5$ the stellar
mass derived from their redshift and multicolor photometry is tightly
correlated to the observed $K$-band magnitude (with a $1\sigma$
dispersion of $\sim 50\%$), at least down to $K=20$. 

$\bullet$ 
The $BzK$ selection and the above correlations ($BzK$ vs. $E(B-V)$ 
and $BzK$ vs. stellar mass) provide a fairly accurate and economic method
that might be statistically applied to the very large samples of
galaxies coming from the current or imminent wide-area surveys 
and/or for galaxy samples beyond the present spectroscopic capabilities.

$\bullet$
A comparison with the UV-selected galaxies at $z\sim
2$ (Steidel et al. 2004), including those at the same $K$ limit, 
shows that $BzK$-selected star-forming galaxies have
typically higher reddening and SFRs. 
Among our $K<20$ sample, the galaxies
satisfying the UGR selection criterion contribute roughly $\sim 15$\%
of the SFRD at $z\sim 2$ produced by the whole K20 sample. On the
other hand the surface density of the Steidel et al. UV-selected galaxies
down to $R=25.5$ is $\sim 10$ times higher than that of $K<20$
star-forming galaxies in the same redshift range, and their
contribution to the SFRD at $z\sim 2$ is a factor $\simgt 2$ higher
than that of the $K<20$, $BzK$-selected galaxies.

$\bullet$
The $BzK$
galaxies at $z\sim 2$ are characterized by a much higher SFR (by a factor
$\sim 8$ on average)
compared to dusty, star-forming EROs ($R-K>5$) at $z\sim 1$,
and $K<20$. We conclude that these vigorous starbursts at $z\sim 2$
are of a different nature compared to highly reddened $z\sim 1$ galaxies.

$\bullet$ 
A $BzK$ analysis of the infrared-selected galaxies with
$J-K>2.3$ (Franx et al. 2003) detected within the K20 survey
shows that those with $z\simgt2$ are likely to be
reddened star-forming objects, rather than passively evolving galaxies. 
A fraction $\sim 50$\% of $J-K>2.3$ galaxies in the K20 survey is
estimated to lie at relatively low redshifts $z\sim1\pm0.5$.
 
$\bullet$ 
$BzK$- and submm/mm-selected galaxies appear to share
properties such as the redshift distribution, high SFRs and high
masses, but the former ones have higher space density while the
latter ones have higher SFRs. An interesting hypothesis is that
submm/mm selected star-forming galaxies might represent extreme subsets of 
$BzK$ galaxies, at least when lying at $1.4\simlt z\simlt2.5$.
 
 $\bullet$ Being
based on the rest-frame shape of the spectra of
 starburst and passive
galaxies, the $BzK$ criterion can be modified to
 select the same
kinds of galaxies within a higher redshift range. In
 this mood, we
propose a $RJL$ criterion to select galaxies within the
 range
$2.5<z<4$, which would complement the $BzK$ selection of
 $1.4<z<2.5$
galaxies.  With uniquely deep $L$-band ($3.6\, \mu$m) data
 that is
becoming available from the Spitzer Space Telescope, this
 criterion
should allow a selection of massive galaxies at $z\sim 3$
that may efficiently complement the traditional LBG selection.
 
$\bullet$
The high masses, SFRs, and metallicities of the bright $BzK$-selected
galaxies at $z\sim 2$, together with a hint for a
strong clustering of them, qualify these galaxies as
possible precursors of $z\sim 1$ passively evolving EROs and $z=0$
early-type galaxies. A $V/V_{\rm max}$ test indicates that
the space density of these galaxies may increase with redshift in the
range $1.4\simlt z\simlt 2.5$.
Current theoretical simulations of hierarchical galaxy formation generally
fail to account simultaneously for the space density of both
passively evolving and star-forming galaxies at $z=2$. Hydrodynamical 
simulations can reproduce our observations better than semyanalitical
models.
 
Some of the above conclusions may be affected by cosmic variance,
given the relatively small size of the explored field. To cope with
this limitation a project is underway to cull $BzK$-selected galaxies
over a $\sim1000$ arcmin$^2$ field, $\sim 20$ times larger than the
full K20 survey area, by combining $K$-band data from ESO telescopes
with optical data from SuprimeCam at the SUBARU telescope (Kong et
al. in preparation), and to follow them up spectroscopically with
VIMOS at the VLT.
The validity of the $BzK$ selection at faint $K>20$
magnitudes will be further tested in great detail with a planned
VLT/FORS2 survey (GMASS project) targeting among others $BzK$
selected galaxies fainter than $K=20$.

\acknowledgments
We are very grateful to Ken Kellermann and John Kelly for having 
provided access to their VLA radio maps of CDFS, and for having measured
radio fluxes for our sources; 
to Piero Rosati, Mario Nonino and the CDFS team for
allowing us to use their $BVRI$ FORS images of the CDFS field; to
Gian Luigi Granato, Rachel Somerville and Kentaro Nagamine
for providing details of their models and for useful
discussions; 
to Micol Bolzonella for the assistence with the {\em hyperz} software;
to Alice Shapley for sending us the
transmission curves of the $U_{\rm n}GR_{\rm s}$ system in a digital form, 
and for discussions.
Finally, we would like to thank the referee, Charles Steidel, 
for constructive comments and suggestions that
resulted in a significant improvement of this paper.
This research was funded in part with an ASI grant (IR-059-02). 
E.D. and A.R. gratefully
acknowledge financial support from the ESO Office for Science. 

\citeindexfalse

\end{document}